

\def\footnoterule{\kern -3pt \hrule width 0truein \kern 2.4pt}
\newcount\notenumber \notenumber=1
\def\note#1{\footnote{$^{\the\notenumber}$}{#1}\global\advance\notenumber by 1}

\font\forteenrmb=cmbx10 at 14pt
\font\twelvermb=cmbx10 at 12pt
\font\elevenrmb=cmbx10 at 11pt
\font\eightrmb=cmbx10 at 8pt

\bigskip
\bigskip
\noindent{\forteenrmb Block Orthogonal Polynomials: II. }

\noindent{\twelvermb Hermite and Laguerre Standard Block Orthogonal Polynomials}
\bigskip
\bigskip
{\leftskip=1.5cm \noindent {\bf Jean-Marie Normand}

\medskip
{\eightrmb
\noindent Service de Physique Th\'eorique, CEA/DSM/SPhT - CNRS/MIPPU/URA 2306

\noindent CEA/Saclay, F-91191 Gif-sur-Yvette Cedex, France

\medskip

\noindent E-mail: jean-marie.normand@cea.fr
}

\bigskip
\bigskip
\noindent{\bf Abstract}

\noindent The standard block orthogonal (SBO) polynomials $P_{i;n}(x),\,0\le i\le n$ are real polynomials
of degree $n$ which are orthogonal with respect to a first Euclidean scalar product to polynomials of degree
less than $i$.
In addition, they are mutually orthogonal with respect to a second Euclidean scalar product.
Applying the general results obtained in a previous paper, we determine and investigate these polynomials
when the first scalar product corresponds to Hermite (resp. Laguerre) polynomials.
These new sets of polynomials, we call Hermite (resp. Laguerre) SBO polynomials,
provide a basis of functional spaces well-suited for some applications requiring to take into account
special linear constraints which can be recast into an Euclidean orthogonality relation.

\par}

\bigskip
\bigskip
\noindent{\elevenrmb 1. Introduction}

\bigskip
\noindent In a first paper [1] (paper I), within the general frame of an Euclidean vector space endowed with
distinct scalar products, and using the Gram-Schmidt orthogonalization (G-SO) process,
we investigate the general properties of {\it standard block orthogonal} (SBO) {\it polynomials}.
By definition, for a given non-negative integer $i$, these real polynomials $P_{i;n}(x),\,n=i,i+1,\ldots$,
of degree $n$, are

\noindent (i) orthogonal to the $i$-dimensional subspace ${\cal P}_i$ of real polynomials of degree
less than $i$ with respect to a first Euclidean scalar product $(\,,\,)$ defined by a non-negative weight
function $w(x)$ on the interval $[a,b]$,
$$(x^m\,,\,P_{i;n}):=\int_a^b dx\,w(x) x^m P_{i;n}(x)=0\quad m=0,\ldots,i-1\quad n=i,i+1,\ldots\,; \eqno(1.1)$$

\noindent (ii) mutually orthogonal with respect to a second Euclidean scalar product $(\,,\,)_2$,
similarly defined by a distinct non-negative weight function $w_2(x)$ on the same interval
\note{More generally, the scalar product $(\,,\,)$ can be defined by the positive measure $\{{\cal D},d\mu\}$
where, $\mu$ is a non-constant and non-decreasing real function on the real domain ${\cal D}$.
Similarly, $(\,,\,)_2$ can be defined by $\{{\cal D}_2,d\mu_2\}$.
See section 3.2 in paper I.},
$$(P_{i;m}\,,\,P_{i;n})_2:=\int_a^b dx\,w_2(x) P_{i;m}(x) P_{i;n}(x)=H_{i;n} \delta_{m,n}
\quad m,n=i,i+1,\ldots\,. \eqno(1.2)$$
In other words, for a given subspace ${\cal P}_i$, there exists a unique {\it orthogonal complement}
${\cal P}_i^\bot$ of codimension $i$, orthogonal with respect to the scalar product $(\,,\,)$,
and complementary with respect to the space of real polynomials ${\cal P}$ of any degree
\note{See equation (3.4) in paper I, written with ${\cal P}_N$ instead of ${\cal P}$ and valid for any non-negative
integer value of $N$.},
$$\bigl({\cal P}_i\,,\,{\cal P}_i^\bot\bigr)=0\quad{\rm and}
\quad{\cal P}={\cal P}_i\oplus{\cal P}_i^\bot\,. \eqno(1.3)$$
Then, $\{P_{i;i},P_{i;i+1},\ldots\}$ is an orthogonal basis of ${\cal P}_i^\bot$ with respect to the scalar
product $(\,,\,)_2$.
These basis polynomials are defined uniquely apart from a constant multiplicative factor for each polynomial.

\smallskip
{\it Classical orthogonal polynomials} (i.e., the Hermite, Laguerre and Jacobi polynomials) are orthogonal
with respect to a scalar product defined with a particular positive measure.
These polynomials arise frequently and have been studied in great detail [2--5]
\note{The standard textbook is [2].
See, also [3] chapter 22, [4] sections 10.6--10.13 or [5] section 8.9.}.
The purpose of this paper is to determine and investigate the SBO polynomials in the special
cases where the first measure, $\{[a,b],w(x)dx\}$ corresponds to the Hermite and then the Laguerre polynomials.
This work has been motivated by the recent consideration of `{\it constrained orthogonal polynomials}' by
Giraud {\it et al} [6--8].
Their physical motivations, sketched in section 1 of paper I, were basically to define orthogonal
basis functions, taking into account special linear constraints (i.e. orthogonality relations),
to provide a functional space well-suited for some specific applications.
Thus, aiming to use the Hohenberg-Kohn variatonal principal [9] for the ground state energy in nuclear physics,
the variable function, i.e. the particle density, has to satisfied the constraint of particle number conservation.
In toy models, expanding the particle density on SBO polynomials, these authors only consider the case $i=1$
(and partly $i=2$) for the Hermite polynomials in [6, 8] and the Laguerre polynomials in [7].
Our works generalize these results to all non-negative integer values of $i$ and provide a systematic study
of the properties of these polynomials, in particular getting new recurrence and differentiation formulae.

This paper is organized as follows.
Explicit expressions and properties of the Hermite (resp. Laguerre) SBO polynomials are given in section 2
(resp. 3), i.e. for:
the connection coefficients in section 2.2 (resp. 3.2),
the differentiation formulae in section 2.3 (resp. 3.3),
the recurrence relations with respect to the degree $n$ in section 2.4 (resp. 3.4)
and with respect to both $i$ and $n$ in section 2.5 (resp. 3.5),
the special value at $x=0$ in section 2.6 (resp. 3.6),
the properties of the zeros in section 2.7 (resp. 3.7),
and finally, the interrelation between the Hermite and the Laguerre cases in section 3.8.
We give our conclusions in section 4.
For completeness and to illustrate how the general methods used allow to recover well-known results
in a unified way, several properties of the Hermite (resp. Laguerre) polynomials are briefly recalled
in appendix A (resp. B).
This aims also at emphasizing the similarities to and the differences from the study of the classical
orthogonal polynomials and the SBO polynomials.
Some expressions related to the Pochammer symbol and generalized hypergeometric series are computed
in appendix C.
Some determinants with entries expressed in terms of gamma function, except possibly for the last row,
are evaluated in appendix D.
Finally, some explicit special cases for the Hermite (resp. Laguerre) SBO polynomials
are listed in appendix E (resp. F).

Throughout the remaining of this paper, the same conventions and notations as in paper I are used
\note{See, [1] at the end of section 1.
In particular, it is recalled that: - {\it monic polynomials}, i.e. with the coefficient of the highest power equal
to one, and also any related quantities are denoted by hatted letters, e.g., $\widehat P_n$
and $\widehat h_n$; - classical orthogonal polynomials are defined according to [2] chapter 22,
[3] chapter X or [4] section 8.9.},
in addition

\noindent -- references to (sub)sections, appendices and equations of paper I are preceded by I-;

\noindent -- $[x]$ denotes the largest integer less than or equal to the real number $x$;

\noindent -- for any function $f:x\mapsto f(x)$, the first (resp. second) derivative with respect to $x$
is denoted by a prime, $f'$ (resp. a double prime, $f''$).

\medskip
\bigskip
\noindent{\elevenrmb 2. Hermite standard block orthogonal polynomials}

\medskip
\noindent Definition and main properties of Hermite polynomials, $H_n,\,n=0,1,\ldots$,
are recalled in appendix A.
They correspond to the measure defined by
$$a:=-\infty\quad b:=\infty\qquad w:=e^{-x^2}\,. \eqno(2.1)$$

\bigskip
\noindent {\it 2.1. Metric tensor components for a particular weight function $w_2$}

\medskip
\noindent {\it (i) Basic quantity $\gamma_{j,k}$.}
With the same interval, $[a,b]$, let us choose,
$$w_2=w_2^{(\mu)}:={\rm e}^{-\mu x^2}\quad\mu>0\,. \eqno(2.2)$$
This choice is motivated by the application considered in [6--8] for the special value $\mu=2$.
Our study will be restricted in (ii) below to this special case, since it allows to push the calculation
to its end analytically.
\smallskip
As shown in section I-3.3.3, $w$ and $w_2$ being even with $a=-b$,
the SBO polynomial $P_{i;n},\,n=i,i+1,\ldots$, as the Hermite polynomials, are even or odd according as $n$
is even or odd, equation I-(3.41),
$$P_{i;n}(-x)=(-1)^n P_{i;n}(x)\,.\eqno(2.3)$$
Actually, the even and odd polynomials separate from each other and one has, equation I-(3.50),
$${\rm if}\ (-1)^{i+n}=1\quad{\rm then}\quad \widehat P_{i-1;n}=\widehat P_{i;n}\quad0\le i-1\le n\,.
\eqno(2.4)$$

From equation I-(3.12), the basic quantity $\gamma_{j,k}(\mu)$ reads
$$\gamma_{j,k}(\mu):=\bigl(H_j\,,\,H_k\bigr)_2=\int_{-\infty}^\infty dx\,e^{-\mu x^2} H_j(x) H_k(x) \eqno(2.5)$$
which vanishes by parity if $j+k$ is odd.
Expanding the Hermite polynomials in terms of powers of $x$ with the coefficients given by equation (A.7)
and using equation (A.4), one finds
$$\eqalignno{
\gamma_{2j,2k}(\mu) &=(-1)^{j+k} 2^{2(j+k)} \Gamma\bigl({\textstyle{1\over2}}+j\bigr)
 \Gamma\bigl({\textstyle{1\over2}}+k\bigr) \cr
&\phantom{=}\times\,\sum_{p=0}^j \sum_{q=0}^k {j\choose p} {k\choose q}
{(-1)^{p+q}\over\Gamma\bigl({\textstyle{1\over2}}+p\bigr) \Gamma\bigl({\textstyle{1\over2}}+q\bigr)}
\,\int_{-\infty}^\infty dx\,e^{-\mu x^2} x^{2(p+q)} \cr
\noalign{\smallskip}
&=(-1)^{j+k} 2^{2(j+k)} \Gamma\bigl({\textstyle{1\over2}}+j\bigr)
\Gamma\bigl({\textstyle{1\over2}}+k\bigr) \mu^{-{1\over2}} \cr
&\phantom{=}\times\,\sum_{p=0}^j \sum_{q=0}^k {j\choose p} {k\choose q}
{\Gamma\bigl({\textstyle{1\over2}}+p+q\bigr)
\over\Gamma\bigl({\textstyle{1\over2}}+p\bigr) \Gamma\bigl({\textstyle{1\over2}}+q\bigr)}
(-\mu)^{-(p+q)}\,. &(2.6)\cr
}$$
The value of the sum over $p$ and $q$ above is considered in lemma C.2 (appendix C).
It can be expressed in terms of an {\it hypergeometric function} or the function $S$ defined by equation
(C.8), then,
$$\gamma_{2j,2k}(\mu)=(-1)^{j+k}\,2^{2(j+k)}\,\Gamma\bigl({\textstyle{1\over2}}+j\bigr)
\,\Gamma\bigl({\textstyle{1\over2}}+k\bigr)\,\mu^{-{1\over2}}
\,S\bigl(j,k,{\textstyle{1\over2}};-\mu^{-1},-\mu^{-1}\bigr)\,. \eqno(2.7)$$
For the odd polynomials, from equation (A.8), a similar calculation yields
$$\gamma_{2j+1,2k+1}(\mu)=(-1)^{j+k}\,2^{2(j+k+1)}\,\Gamma\bigl({\textstyle{3\over2}}+j\bigr)
\,\Gamma\bigl({\textstyle{3\over2}}+k\bigr)\,\mu^{-{3\over2}}
\,S\bigl(j,k,{\textstyle{3\over2}};-\mu^{-1},-\mu^{-1}\bigr)\,. \eqno(2.8)$$
Alternately, these last two equations can be obtained using, with some care, the definite integrals
given in [5, 10]
\note{See, [5] equation 7.374 5 where, the second argument of the hypergeometric function
should be $-n$ instead of $n$ (the result has to be symmetric in $m$ and $n$) and in addition for $\alpha^2$ real,
$\alpha$ has to be taken positive.
See also [10] 2.20.16 1 where, the factor $(\pm1)^{m+n}$ should be replaced by $(-1)^{m+n}$,
or else, 2.20.16 6.
In both cases, one has to use also the linear transformation formula [3] 15.3.6.}.
Another way to get the basic quantity $\gamma_{j,k}(\mu)$ is to compute the scalar product
$(F_s\,,\,F_t)_2$ where, $F_z(x)$ is the generating function of Hermite polynomials given by equation (A.29).
Then, $\gamma_{j,k}(\mu)$ is related to the coefficient of $s^j t^k$ in the expansion of this scalar product,
i.e.,
$$\eqalignno{
(F_s\,,\,F_t)_2 &=\int_{-\infty}^\infty dx\;e^{-\mu x^2} F_s(x) F_t(x)
=\sum_{j,k=0}^\infty {1\over j!\,k!}\,\gamma_{j,k}(\mu) s^j t^k \cr
&=\Bigl({\pi\over\mu}\Bigr)^{1\over2} \exp{2st+(1-\mu)(s^2+t^2)\over\mu}\,. &(2.9)\cr
}$$

For $\mu=1$, i.e. for $w_2=w$, one has to recover the orthogonality condition of Hermite
polynomials.
Indeed, from equation (C.14),
$$S\bigl(j,k,{\textstyle{1\over2}};-1,-1\bigr)={j!\over\Gamma\bigl({\textstyle{1\over2}}+j\bigr)}
\delta_{j,k} \eqno(2.10)$$
whence, with the {\it duplication formula} [3]
\note{See, e.g., [3] 6.1.18,
$\Gamma(z) \Gamma\bigl({\textstyle{1\over2}}+z\bigr)=2^{-2z+1} \pi^{1\over2} \Gamma(2z)\,.$},
$$\gamma_{2j,2k}(1)=2^{4j} j!\,\Gamma\bigl({\textstyle{1\over2}}+j\bigr)\,\delta_{j,k}
=2^{2j} (2j)!\,\pi^{1\over2} \delta_{j,k} \eqno(2.11)$$
recovering thereby the value of $h_{2j}$ given by equation (A.6).
The calculation is similar for $\gamma_{2j+1,2k+1}(1)$.
The results follows readily also from equation (2.9), indeed,
$(F_s\,,\,F_t)_2=\pi^{1\over2} e^{2st}=\sum_{j=0}^\infty h_j (st)^j/(j!)^2$.

\medskip
\noindent {\it (ii) The special case $w_2=w_2^{(2)}=e^{-2x^2}$.}
When $\mu=2$, one gets simple expressions.
Indeed, from equation (C.15) setting $z=-{1\over2}$, one has
$$\eqalignno{
\gamma_{2j,2k}(2) &=(-1)^{j+k}\,2^{j+k-{1\over2}}
\,\Gamma\bigl({\textstyle{1\over2}}+j+k\bigr) &(2.12)\cr
\noalign{\smallskip}
\gamma_{2j+1,2k+1}(2) &=(-1)^{j+k}\,2^{j+k+{1\over2}}
\,\Gamma\bigl({\textstyle{3\over2}}+j+k\bigr) &(2.13)\cr
}$$
or, equivalently [5]
\note{Equation (2.12) is given in [5] equation 7.374 2, with the wrong sign:
the result should be symmetric in $j$ and $k$ and positive for $j=k$.},
$$\gamma_{j,k}(2)=\cases{
(-1)^{[j/2]+[k/2]}\,2^{{1\over2}(j+k-1)}\,\Gamma\bigl({\textstyle{1\over2}}(j+k+1)\bigr) &if $j+k$ even\cr
\noalign{\smallskip}
0 &otherwise.\cr
} \eqno(2.14)$$
Alternately, this result follows also from equation (2.9),
$$(F_s\,,\,F_t)_2=\Bigl({\pi\over2}\Bigr)^{1\over2} e^{-{1\over2}(s-t)^2}=\Bigl({\pi\over2}\Bigr)^{1\over2}
\sum_{n=0}^\infty \sum_{m=0}^{2n} (-1)^{n-m} 2^{-n} {(2n)!\over n!\,m!\,(2n-m)!}\,s^m t^{2n-m}\,.
\eqno(2.15)$$
Setting $j=m$, $k=2n-m$ and using the duplication formula to get
$(2n)!/n!=\pi^{-{1\over2}} 2^{2n} \Gamma({1\over2}+n)$, one recovers equation (2.14).

\medskip
\noindent {\it From now on, except explicitly noted, only this special case $\mu=2$ will be considered.}

\noindent Then, the corresponding polynomials $P_{i;n},\,0\le i\le n$ are called the {\it Hermite SBO polynomials}.
The case $i=1$ has already been considered in [6, 8] with different choices of weight functions and normalization.
(See footnote 50 in appendix E for interrelations with our polynomial $\widehat P_{1;n}$.)
The case $i=2$, only for the odd polynomials, has been also considered in [6], as already discussed
in the conclusion of paper I (see footnote 44 of paper I).
(See footnote 51 for interrelations with our polynomial $\widehat P_{2;2n+1}$.)

\bigskip
\noindent {\it 2.2. Connection coefficients between block orthogonal and classical polynomials}

\medskip
\noindent The relevant quantities are given by equations I-(3.42)--I-(3.47) in terms of determinants with the
$(j,k)$-element of the type $\Gamma(c+j+k)$, except possibly for the last row.
This kind of determinant is evaluated in lemma D (appendix D).
From equations I-(3.43), I-(3.46), (2.12), (2.13) and (D.3), one finds for $0\le i\le n$,
$$\eqalignno{
Z_{i;n}^{({\rm e})} &=2^{(i+n-{1\over2})(n-i+1)} \prod_{j=i}^n (j-i)!
\,\Gamma\bigl({\textstyle{1\over2}}+i+j\bigr)\qquad Z_{i;i-1}^{({\rm e})}:=1 &(2.16)\cr
Z_{i;n}^{({\rm o})} &=2^{(i+n+{1\over2})(n-i+1)} \prod_{j=i}^n (j-i)!
\,\Gamma\bigl({\textstyle{3\over2}}+i+j\bigr)\qquad Z_{i;i-1}^{({\rm o})}:=1 &(2.17)\cr
}$$
and therefore, from equations I-(3.44) and I-(3.47) with $k_n$ given by equation (A.3) [6]
\note{These equations generalize the formulae given in [6] (4) (resp. (19)) for the special case $i=1$
(resp. $i=2$ for the odd polynomials) with different choices of $w$, $w_2$ and normalization,
see equation (E.12) and footnote 50 (resp. 51)},
$$\eqalignno{
\widehat H_{2i-1;2n}=\widehat H_{2i;2n} &=2^{-(2n+{1\over2})} (n-i)!
\,\Gamma\bigl({\textstyle{1\over2}}+i+n\bigr) &(2.18)\cr
\noalign{\smallskip}
\widehat H_{2i;2n+1}=\widehat H_{2i+1;2n+1} &=2^{-(2n+{3\over2})} (n-i)!
\,\Gamma\bigl({\textstyle{3\over2}}+i+n\bigr)\,. &(2.19)\cr
}$$
The monic SBO polynomials follow from equations I-(3.42), I-(3.45) and (D.2), for $0\le i\le n$,
$$\eqalignno{
\widehat P_{2i-1;2n}=\widehat P_{2i;2n} &=\sum_{m=i}^n 2^{-(n+m)} {n-i\choose m-i}
{\Gamma\bigl({1\over2}+i+n\bigr)\over\Gamma\bigl({1\over2}+i+m\bigr)} H_{2m} &(2.20)\cr
\noalign{\smallskip}
\widehat P_{2i;2n+1}=\widehat P_{2i+1;2n+1} &=\sum_{m=i}^n 2^{-(1+n+m)} {n-i\choose m-i}
{\Gamma\bigl({3\over2}+i+n\bigr)\over\Gamma\bigl({3\over2}+i+m\bigr)} H_{2m+1} &(2.21)\cr
}$$
where, according to equations I-(3.13), (2.4) and (A.3), for $i\ge0$,
$$\widehat P_{i;i}=\widehat H_i:=2^{-i}\,H_i
\qquad\widehat P_{i;i+1}=\widehat H_{i+1}=2^{-(i+1)}\,H_{i+1}\,. \eqno(2.22)$$
The orthogonality and normalization of these polynomials with respect to the scalar product
$(\,,\,)_2$ can be checked directly using equation (C.14), e.g., for the even polynomials with equations (2.20)
and (2.14),
$$\eqalignno{
&\int_{-\infty}^\infty dx\,e^{-2x^2} \widehat P_{2i;2n}(x) \widehat P_{2i;2n'}(x) \cr
&=2^{-(n+{n'}+{1\over2})} \Gamma\bigl({\textstyle{1\over2}}+i+n\bigr)
\Gamma\bigl({\textstyle{1\over2}}+i+n'\bigr) S(n-i,n'-i,{\textstyle{1\over2}}+2i;-1;-1) \cr
\noalign{\smallskip}
&=2^{-(2n+{1\over2})} (n-i)!\,\Gamma\bigl({\textstyle{1\over2}}+i+n\bigr) \delta_{n,{n'}} &(2.23)\cr
}$$
recovering equation (2.18).

Equations (2.20) and (2.21) can be inverted to give for $0\le i\le n$,
$$\eqalignno{
H_{2n} &=\sum_{m=i}^n (-1)^{n-m} 2^{n+m} {n-i\choose m-i}
{\Gamma\bigl({1\over2}+i+n\bigr)\over\Gamma\bigl({1\over2}+i+m\bigr)} \widehat P_{2i;2m} &(2.24)\cr
\noalign{\smallskip}
H_{2n+1} &=\sum_{m=i}^n (-1)^{n-m} 2^{1+n+m} {n-i\choose m-i}
{\Gamma\bigl({3\over2}+i+n\bigr)\over\Gamma\bigl({3\over2}+i+m\bigr)}\widehat P_{2i+1;2m+1}\,. &(2.25)\cr
}$$
As for the Hermite polynomials in appendix A,
these relations can be obtained either using the general formulae I-(A.22) and I-(A.28)
where, the determinants which occur can be evaluated with equation (D.2), or directly,
using lemma A (appendix A), e.g., for the even polynomials, setting $j=m-i$, $k=n-i$,
$\rho_j=2^{-(i+j)} \Gamma({1\over2}+2i+j)$ and $\sigma_j=(-1)^j 2^{i+j} \Gamma({1\over2}+2i+j)$.

It should be noted that the connection coefficients relating the polynomials $\widehat P_{i;n}$ and $H_n$
in equations (2.20), (2.21), (2.24) and (2.25) have the same structure as the connection coefficients relating
$H_n$ and the monomials in equations (A.7), (A.8), (A.12) and (A.13), respectively.
This is due to the fact that in both cases, the metric tensor components are of the
same type $\Gamma(c+j+k)$, apart from some factors depending separately on the indices $j$ and $k$,
see equations (2.14) and (A.4).
(See also equation (2.30) below.)

As already noted in section 3.2 of paper I (see equations I-(3.18) and I-(3.19)),
for $i=0$, $\widehat P_{0;n}$ has to be the unique standard monic orthogonal polynomial $\widehat Q_{2;n}$
corresponding to the weight function $w_2$.
Since $w_2^{(2)}(x)=w(\sqrt2 x)$, with the normalization (A.3) and changing $x$ into $\sqrt2 x$ in the integral
defining the scalar product of Hermite polynomials, one gets for $n\ge0$,
$$\widehat P_{0;n}(x)=2^{-{1\over2}n} \widehat H_n\bigl(\sqrt2 x\bigr)
=2^{-{3\over2}n} H_n\bigl(\sqrt2 x\bigr)\,. \eqno(2.26)$$
Thereby, setting $i=0$ in equations (2.20), (2.21), (2.24) and (2.25), using the duplication formula and
changing $m$ into $n-m$ in the sum over m yield the connection coefficients between the two basic sets
of polynomials $\{H_m,\,m=0,\ldots,n\}$ and $\{H_m(\sqrt2 x),\,m=0,\ldots,n\}$ for $n\ge0$,
$$\eqalignno{
H_n\bigl(\sqrt2 x\bigr) &=\sum_{m=0}^{[n/2]} 2^{{1\over2}(n-2m)} {n!\over m!\,(n-2m)!} H_{n-2m}(x) &(2.27)\cr
H_n(x) &=\sum_{m=0}^{[n/2]} (-1)^m 2^{-{1\over2}n} {n!\over m!\,(n-2m)!}
H_{n-2m}\bigl(\sqrt2 x\bigr)\,. &(2.28)\cr
}$$
These remarkable relations can be checked using the integral [5, 10]
\note{See [5] equation 7.374 4 or [10] 2.20.16.7. p 503.}
and the orthogonality relation of Hermite polynomials with $h_n$
given by equation (A.6), e.g., one has
$$\int_{-\infty}^\infty dx\,e^{-x^2} H_n(\sqrt2 x) H_{n-2m}(x)
=2^{{3\over2}(n-2m)} {n!\over m!} \pi^{1\over2}
=2^{{1\over2}(n-2m)} {n!\over m!\,(n-2m)!} h_{n-2m}\,. \eqno(2.29)$$
Another way to get the relation (2.27) is as follows.
From equations (A.4) and (2.14), one has
$$(x^j\,,\,x^k)=(-1)^{[j/2]+[k/2]} 2^{-{1\over2}(j+k-1)} \gamma_{j,k}(2)
=\sqrt2 \Bigl((-1)^{[j/2]} 2^{-{1\over2}j} H_j\,,\,(-1)^{[k/2]} 2^{-{1\over2}k} H_k\Bigr)_2 \eqno(2.30)$$
namely, the metric tensor components of the scalar product $(\,,\,)$ in the basis $\{x^n,\,n=0,1,\ldots\}$
and the metric tensor components of the scalar product $(\,,\,)_2$ in the basis
$\{(-1)^{[n/2]} 2^{-{1\over2}n} H_n,\,n=0,1,\ldots\}$ are equal apart from a constant factor.
Hence, while G-SO with respect to $(\,,\,)$ of the basis $\{x^n,\,n=0,1,\ldots\}$ defines
the Hermite polynomial $H_n$, G-SO with respect to $(\,,\,)_2$ of the basis
$\{(-1)^{[n/2]} 2^{-{1\over2}n} H_n,\,n=0,1,\ldots\}$ defines the polynomial $\widehat P_{0;n}$.
In other words, the linear mapping defined by $x^n\,\mapsto\,(-1)^{[n/2]} 2^{-{1\over2}n} H_n,\,n=0,1,\ldots$,
is such that $H_n\,\mapsto\,\propto\widehat P_{0,n},\ n=0,1,\ldots$.
Therefore, apart from a multiplicative factor $c$, $\widehat P_{0,n}$ reads as a linear combination of
$(-1)^{[m/2]} 2^{-{1\over2}m} H_m,\,m=0,1,\ldots$ with {\it the same connection coefficients} as in the
expansion of $H_n$ in terms of the monomials
\note{It is underlined in section I-2.1 that all the relevant quantities in G-SO procedure depend only
on the metric tensor components.},
i.e. from equation (A.9),
$$\widehat P_{0;n}=c \sum_{m=0}^{[n/2]} {(-1)^m 2^{n-2m}\over m!\,(n-2m)!}
\Bigl((-1)^{[(n-2m)/2]} 2^{-{1\over2}(n-2m)} H_{n-2m}\Bigr)\,. \eqno(2.31)$$
With $\widehat P_{0;n}$ a monic polynomial, computing the coefficient of $x^n$ fixes
$c=(-1)^{[n/2]} 2^{-{3\over2}n} n!$, recovering thereby equation (2.27) with (2.26).

From equations (2.4) and (2.26), one has
$$\widehat P_{1;2n+1}(x)=\widehat P_{0;2n+1}(x)=2^{-3(n+{1\over2})} H_{2n+1}(\sqrt2 x) \eqno(2.32)$$
(see also the comments about equation I-(3.52) and in the conclusion of paper I,
$\widehat P_{1;2n+1}=\widehat Q_{2;2n+1}$).
This relation can be checked setting $i=0$ in equation (2.21) and comparing with the expansion
of $H_{2n+1}(\sqrt2 x)$ given by equation (2.26).

It follows from equations I-(3.10), I-(3.11), (2.20), (2.21) and (A.10) that,
$$\eqalignno{
\widehat R_{i;n} &=0 &(2.33)\cr
\widehat S_{2i-1;2n}=\widehat S_{2i;2n} &=-{1\over4} \Bigl(i(2i-1)+n(2n-1)\Bigr) &(2.34)\cr
\widehat S_{2i;2n+1}=\widehat S_{2i+1;2n+1} &=-{1\over4} \Bigl(i(2i+1)+n(2n+1)\Bigr)\,. &(2.35)\cr
}$$
For $i=0$, one gets from above $\widehat S_{0;n}=-n(n-1)/8$, which can be checked using equations
(2.26) and (A.10).

For the general properties of SBO polynomials about the {\it projection operators}, the {\it integral
representations} and the {\it normalizations}, see sections 3.3.1, 3.3.2 and 3.3.4 in paper I, respectively.

\bigskip
\noindent {\it 2.3. Differentiation formulae}

\medskip
\noindent Let us summarize well-known results for all classical orthogonal polynomials:

\noindent $\bullet$ as for any standard orthogonal polynomials, the {\it three term recurrence formula} I-(3.55) holds;

\noindent $\bullet$ the weight function which defines the scalar product satisfies a differential equation.
This implies several differentiation formulae, e.g., the {\it linear homogeneous differential
equation of the second order}
\note{See, e.g., [4] 10.6 (ii) p 164.},
$$A(x)\,y''+B(x)\,y'+\lambda_n\,y=0 \eqno(2.36)$$
where, $A(x)$ and $B(x)$ are independent of $n$ and $\lambda_n$ is independent of $x$.

\smallskip
All these relations follow from the three term recurrence formula and one of the first order
differentiation formulae.
Moreover, among the three formulae---the recurrence equation, a first order differentiation equation and the
second order differential equation---any pair of equations implies the third one.
In particular, two first order differentiation formulae or, the first and second order differentiation formulae,
imply the recurrence formula.
This is the method used to get a five term recurrence formula in the next subsection.
However, while the differentiation formulae for classical orthogonal polynomials can be derived in two ways,
as recalled in appendices A (item 4) and B (item 4), the second method only can be used for block orthogonal polynomials.
This is due to the fact, already encountered in sections 3.3.5 and 3.3.7 of paper I,
that the scalar product $(\,,\,)_0$, defined by equation I-(3.5)
\note{Any Euclidean scalar product $(\,,\,)_1$ defined on ${\cal P}_i$, together with $(\,,\,)_2$ defined on
${\cal P}_i^\bot$ induce a new Euclidean structure on ${\cal P}$ with the scalar product defined by
$(p\,,\,q)_0:=(p_i\,,\,q_i)_1+(p_i^\bot\,,\,q_i^\bot)_2$ for any
$p=p_i+p_i^\bot\in{\cal P},\ p_i\in{\cal P}_i,\ p_i^\bot\in{\cal P}_i^\bot$
and $q=q_i+q_i^\bot\in{\cal P},\ q_i\in{\cal P}_i,\ q_i^\bot\in{\cal P}_i^\bot$.},
is no longer defined by a single integral.
Here, for a given nonzero $i$, one has to deal with both scalar products $(\,,\,)$ and $(\,,\,)_2$,
each of them defined by an integral with a different weight function.

For the scalar product $(\,,\,)$, the weight function $w$ satisfies the differential equation (A.17),
which implies the relations (A.18) and (A.19).
Similarly, for the scalar product $(\,,\,)_2$ associated with the weight function $w_2^{(\mu)}$
defined by equation (2.2), one has for any polynomials $f$ and $g$,
$$\eqalignno{
w_2^{(\mu)}{}' &=-2\mu x w_2^{(\mu)} &(2.37)\cr
\noalign{\smallskip}
(f\,,\,g')_2 &=-(f'-2\mu x f\,,\,g)_2 &(2.38)\cr
\noalign{\smallskip}
(f''-2\mu x f'\,,\,g)_2 &=-(f'\,,\,g')_2=(f\,,\,g''-2\mu x g')_2\,. &(2.39)\cr
}$$
\noindent (i) For $0\le\ell<i$, the polynomial $H'_\ell$ of degree $\ell-1$ is in the subspace ${\cal P}_i$.
Therefore, from the orthogonality condition (1.1), $H'_\ell$ and $\widehat P_{i;n}$ for $i\le n$ are orthogonal
with respect to the scalar product $(\,,\,)$, and from equation (A.18), one has
\note{Actually, equation (2.40) still holds for $\ell=i$. Since from equation (2.22),
$\widehat P_{i;i}=\widehat H_i$, the component $\alpha_{i;n,i}$ considered in equation (2.41) below, vanishes.
This result is obtained also by parity.},
$$0=\bigl(\widehat P_{i;n}\,,\,H'_\ell\bigr)
=-\bigl(\widehat P'_{i;n}-2x \widehat P_{i;n}\,,\,H_\ell\bigr)\,. \eqno(2.40)$$
Hence, the particular polynomial $\widehat P'_{i;n}-2x \widehat P_{i;n}$ of degree $n+1$ is orthogonal to
the subspace ${\cal P}_i$ with respect to the scalar product $(\,,\,)$, i.e. it belongs to the
orthogonal subspace ${\cal P}_i^\bot$, see equation (1.3),  and thus reads
$\sum_{j=i}^{n+1} \alpha_{i;n,j} \widehat P_{i;j}$.
Computing the coefficient of $x^{n+1}$ in $\widehat P'_{i;n}-2x \widehat P_{i;n}$ fixes $\alpha_{i;n,n+1}=-2$.
By parity, $\alpha_{i;n,n},\alpha_{i;n,n-2},\ldots$ vanish.
Now, with respect to the scalar product $(\,,\,)_2$, one has from equations (1.2) and (2.38),
for $j=i,\ldots,n+1$,
$$\eqalignno{
\alpha_{i;n,j} \widehat H_{i;j} &=\bigl(\widehat P_{i;j}\,,\,\widehat P'_{i;n}-2x \widehat P_{i;n}\bigr)_2
=\bigl(\widehat P_{i;j}\,,\,\widehat P'_{i;n}\bigr)_2-\bigl(2x \widehat P_{i;j}\,,\,\widehat P_{i;n}\bigr)_2 \cr
\noalign{\smallskip}
&=-\bigl(\widehat P'_{i;j}-2(\mu-1) x \widehat P_{i;j}\,,\,\widehat P_{i;n}\bigr)_2\,. &(2.41)\cr
}$$
In the special case $\mu=2$, it turns out that $\widehat P'_{i;j}-2(\mu-1) x \widehat P_{i;j}$
is the particular polynomial $\widehat P'_{i;j}-2x \widehat P_{i;j}$, introduced above for $j=n$
(i.e. of degree $j+1$ in the subspace ${\cal P}_i^\bot$), and which reads
$\sum_{k=i}^{j+1} \alpha_{i;j,k} \widehat P_{i;k}$.
Thus, for $\mu=2$, {\it considered only henceforth}, one has for $i\le j\le n+1$,
$$\alpha_{i;n,j} \widehat H_{i;j}=-\alpha_{i;j,n} \widehat H_{i;n} \sum_{k=i}^{j+1} \delta_{k,n}\,.
\eqno(2.42)$$
For $n=i$, all the $\alpha$-coefficients are known: $\alpha_{i;i,i}=0,\,\alpha_{i;i,i+1}=-2$.
For $n\ge i+1$, it follows from above that for $n\ge i+2$ and $j=i,\ldots,n-2$, $\alpha_{i;n,j}$ vanishes.
It remains to consider the case $j=n-1$.
With $\alpha_{i;n-1,n}=-2$, equations (2.18) and (2.19), one gets
(by inspection of the four cases according to the parity of $i$ and $n$),
$$\kappa_{i;n}:=\alpha_{i;n,n-1}=2 {\widehat H_{i;n}\over\widehat H_{i;n-1}}
={1\over2} \bigl(n-(-1)^{i+n}i\bigr)\,. \eqno(2.43)$$
Finally, for $\mu=2$ and $0\le i\le n$, one finds the {\it first order differentiation formula}
\note{This generalizes the formula given in [6] (7) for the special case $i=1$ with different choices of $w$ and
$w_2$, see footnote 50.},
$$\widehat P'_{i;n}=-2 \widehat P_{i;n+1}+2x \widehat P_{i;n}
+\kappa_{i;n} \widehat P_{i;n-1} \eqno(2.44)$$
which still holds for $n=i$, since then $\kappa_{i;n}$ vanishes.
This equation has to be compared with equation (A.26).
For $n=i$, both equations coincide, using equation (2.22).
For $i=0$, using equation (2.26), one gets
$$H'_n=-{1\over2} H_{n+1}+x H_n+n H_{n-1} \eqno(2.45)$$
which follows from, e.g., the differentiation formula (A.21) and the recurrence formula (A.16).

With the boundary value of $\widehat P_{i;i}$, given in terms of Hermite polynomials by equation (2.22),
the differentiation formula (2.44) can be used to get step by step $\widehat P_{i;n}$ for all $i$ and $n$
with $0\le i\le n$.
One can also take advantage of the parity relation (2.4) to divide by two the number of polynomials to be computed.
A table of polynomials $\widehat P_{i;n}$ for $i=0,1,2$ and $n=i,i+1,\ldots,i+4$ is given in appendix E.

Following the same steps as above, for $0\le\ell<i\le n$, the polynomial $xH'_\ell$ of degree $\ell$
is orthogonal to $\widehat P_{i;n}$ with respect to the scalar product $(\,,\,)$,
and from equation (A.18), one has
$$0=\bigl(\widehat P_{i;n}\,,\,xH'_\ell\bigr)=\bigl(x\widehat P_{i;n}\,,\,H'_\ell\bigr)
=-\bigl(x\widehat P'_{i;n}+\widehat P_{i;n}-2x^2\widehat P_{i;n}\,,\,H_\ell\bigr)
=-\bigl(x\widehat P'_{i;n}-2x^2\widehat P_{i;n}\,,\,H_\ell\bigr)\,. \eqno(2.46)$$
Hence, the particular polynomial $x\widehat P'_{i;n}-2\,x^2\widehat P_{i;n}$ of degree $n+2$ belongs to
${\cal P}_i^\bot$  and reads $\sum_{j=i}^{n+2}\beta_{i;n,j}\,\widehat P_{i;j}$.
Computing the coefficient of $x^{n+2}$ in $x\widehat P'_{i;n}-2x^2\widehat P_{i;n}$ fixes $\beta_{i;n,n+2}=-2$.
By parity, $\beta_{i;n,n+1},\beta_{i;n,n-1},\ldots$ vanish.
Now, with respect to the scalar product $(\,,\,)_2$, one has from equations (1.2) and (2.38), for $i\le j\le n+1$,
$$\eqalignno{
\beta_{i;n,j}\,\widehat H_{i;j} &=\bigl(\widehat P_{i;j}\,,\,x\widehat P'_{i;n}-2x^2\widehat P_{i;n}\,)_2
=\bigl(x\widehat P_{i;j}\,,\,\widehat P'_{i;n}\bigr)_2
-\bigl(2x^2\widehat P_{i;j}\,,\,\widehat P_{i;n}\bigr)_2 \cr
\noalign{\smallskip}
&=-\bigl(x\widehat P'_{i;j}-2(\mu-1)x^2\widehat P_{i;j}\,,\,\widehat P_{i;n}\bigr)_2
-\widehat H_{i;n} \delta_{j,n}\,. &(2.47)\cr
}$$
In the special case $\mu=2$, it turns out that
$x\widehat P'_{i;j}-2(\mu-1)x^2\widehat P_{i;j}$ is the particular polynomial
$x\widehat P'_{i;j}-2 x^2\widehat P_{i;j}$, introduced above for $j=n$
(i.e. of degree $j+2$ in the subspace ${\cal P}_i^\bot$), and which reads
$\sum_{k=i}^{j+2} \beta_{i;j,k} \widehat P_{i;k}$.
Thus, one has for $i\le j\le n+1$,
$$\beta_{i;n,j} \widehat H_{i;j}=\Bigl(-\beta_{i;j,n} \sum_{k=i}^{j+2} \delta_{k,n}
-\delta_{j,n}\Bigr) \widehat H_{i;n} \eqno(2.48)$$
and therefore, $\beta_{i;n,n}=-{1\over2}$.
Then, for $n=i$ and $n=i+1$, all the $\beta$-coefficients are known:
$\beta_{i;i,i}=-{1\over2},\,\beta_{i;i,i+1}=0,\,\beta_{i;i,i+2}=-2$ and
$\beta_{i;i+1,i}=0,\,\beta_{i;i+1,i+1}=-{1\over2},\,\beta_{i;i+1,i+2}=0,\,\beta_{i;i+1,i+3}=-2$.
For $n\ge i+2$, it follows from equation (2.48) that for $n\ge i+3$ and $j=i,\ldots,n-3$, $\beta_{i;n,j}$ vanishes.
It remains to consider the case $j=n-2$.
From $\beta_{i;n-2,n}=-2$ and equation (2.43), one gets
$$\beta_{i;n,n-2}=2 {\widehat H_{i;n}\over\widehat H_{i;n-2}}
={1\over2} \kappa_{i;n} \kappa_{i;n-1}\,. \eqno(2.49)$$
Finally, in the case $\mu=2$, for $0\le i\le n$ and $\kappa_{i;n}$ given by equation (2.43),
one finds another {\it first order differentiation formula},
$$x\widehat P'_{i;n}=-2 \widehat P_{i;n+2}+{1\over2} (4x^2-1) \widehat P_{i;n}
+{1\over2} \kappa_{i;n} \kappa_{i;n-1} \widehat P_{i;n-2} \eqno(2.50)$$
which still holds for $n=i$ and $n=i+1$, since then $\kappa_{i;n} \kappa_{i;n-1}$ vanishes.
This equation, which relates polynomials having the same parity, has to be compared with equation (A.27).
For $i=0$, using equation (2.26), one gets
$$4 xH'_n=-H_{n+2}+2(2x^2-1) H_n+4n(n-1) H_{n-2} \eqno(2.51)$$
which follows, e.g., from the differentiation formula (A.27) and the recurrence formula (A.28).

Starting with the values of $\widehat P_{i;i}$ and $\widehat P_{i;i+1}$ given by equation (2.22),
the differentiation formula (2.50) allows, as equation (2.44), to compute step by step $\widehat P_{i;n}$
for all $i$ and $n$ with $0\le i\le n$ (see appendix E).

\smallskip
\noindent (ii) Following the same arguments again,
for $0\le\ell<i\le n$, the polynomial $H''_\ell-2xH'_\ell$ of degree $\ell$ is orthogonal to $\widehat P_{i;n}$
with respect to the scalar product $(\,,\,)$, and from equation (A.19), one has
$$0=\bigl(\widehat P_{i;n}\,,\,H''_\ell-2xH'_\ell\bigr)
=\bigl(\widehat P''_{i;n}-2x\widehat P'_{i;n}\,,\,H_\ell\bigr)\,. \eqno(2.52)$$
Hence, the particular polynomial $\widehat P''_{i;n}-2\,x\widehat P'_{i;n}$ of degree $n$ belongs to
${\cal P}_i^\bot$  and reads $\sum_{j=i}^n \gamma_{i;n,j} \widehat P_{i;j}$.
Computing the coefficient of $x^{n}$ in $\widehat P''_{i;n}-2x\widehat P'_{i;n}$ fixes $\gamma_{i;n,n}=-2n$.
By parity, $\gamma_{i;n,n-1},\gamma_{i;n,n-3},\ldots$ vanish.
With respect to the scalar product $(\,,\,)_2$, one has from equations (1.2), (2.38) and (2.39),
for $i\le j\le n$,
$$\eqalignno{
\gamma_{i;n,j}\,\widehat H_{i;j} &=\bigl(\widehat P_{i;j}\,,\,\widehat P''_{i;n}-2x\widehat P'_{i;n}\bigr)_2
=\bigl(\widehat P_{i;j}\,,\,\widehat P''_{i;n}-2\mu x\widehat P'_{i;n}\bigr)_2
+2(\mu-1)\bigl(x\widehat P_{i;j}\,,\,\widehat P'_{i;n}\bigr)_2 \cr
\noalign{\smallskip}
&=\Bigl(\bigl(\widehat P''_{i;j}-2(\mu-1) x\widehat P'_{i;j}\bigr)
-2\mu\bigl(x\widehat P'_{i;j}-2(\mu-1) x^2\widehat P_{i;j}\bigr)\,,\,\widehat P_{i;n}\Bigr)_2
-2(\mu-1)\widehat H_{i;n} \delta_{j,n}\,. &(2.53)\cr
}$$
In the special case $\mu=2$, once again,
$\widehat P''_{i;j}-2(\mu-1) x\widehat P'_{i;j}$ (resp. $x\widehat P'_{i;j}-2(\mu-1)x^2\widehat P_{i;j}$)
is the particular polynomial
$\widehat P''_{i;j}-2x\widehat P'_{i;j}$ (resp. $x\widehat P'_{i;j}-2x^2\widehat P_{i;j}$),
introduced above for $j=n$
(i.e. of degree $j$ (resp. $j+2$) in the subspace ${\cal P}_i^\bot$), and which reads
$\sum_{k=i}^j \gamma_{i;j,k} \widehat P_{i;k}$ (resp. $\sum_{k=i}^{j+2} \beta_{i;j,k} \widehat P_{i;k})$.
Thus, one has for $i\le j\le n$,
$$\gamma_{i;n,j} \widehat H_{i;j}=\Bigl(\gamma_{i;j,n}\,\sum_{k=i}^j \delta_{k,n}-4\beta_{i;j,n}
\sum_{k=i}^{j+2} \delta_{k,n}-2 \delta_{j,n}\Bigr) \widehat H_{i;n}\,.\eqno(2.54)$$
For $n=i$ and $n=i+1$, all the $\gamma$-coefficients are known: $\gamma_{i;i,i}=-2i$ and
$\gamma_{i;i+1,i}=0,\,\gamma_{i;i+1,i+1}=-2(i+1)$.
For $n\ge i+2$, it follows from above that for $n\ge i+3$ and $j=i,\ldots,n-3$, $\gamma_{i;n,j}$ vanishes.
It remains to consider the case $j=n-2$.
From $\beta_{i;n-2,n}=-2$ and  $\beta_{i;n,n}=-{1\over2}$, one gets
$$\gamma_{i;n,n-2}=-4\beta_{i;n-2,n} {\widehat H_{i;n}\over\widehat H_{i;n-2}}
=2 \kappa_{i;n} \kappa_{i;n-1}\,. \eqno(2.55)$$
Finally, in the case $\mu=2$, for $0\le i\le n$ and $\kappa_{i;n}$ given by equation (2.43),
one finds the {\it second order differentiation formula}
\note{This generalizes the formula given in [6] (15) for the special case $i=1$ with different choices
of $w$ and $w_2$, see footnote 50.},
$$\widehat P''_{i;n}-2x\widehat P'_{i;n}+2n\widehat P_{i;n}
=2\kappa_{i;n} \kappa_{i;n-1}\,\widehat P_{i;n-2} \eqno(2.56)$$
which still holds for $n=i$ and $n=i+1$, since then $\kappa_{i;n} \kappa_{i;n-1}$ vanishes.
This equation has to be compared with the differential equation (A.24).
For $n=i$ and $n=i+1$, both equations coincide, using equation (2.22).
For $i=0$, from equation (2.26), one gets
$$H''_n-x\,H'_n+n\,H_n=2n(n-1)\,H_{n-2} \eqno(2.57)$$
which follows, e.g., from equations (A.22) and (A.24).

\bigskip
\noindent {\it 2.4. Recurrence relation with respect to the degree $n$}

\medskip
\noindent In the case $\mu=2$, eliminating $\widehat P'_{i;n}$ between the two first order differentiation
formulae (2.44) and (2.50) yields for $0\le i\le n$ the {\it five term linear recurrence formula},
$$\widehat P_{i;n+2}-x\widehat P_{i;n+1}+{1\over4}\widehat P_{i;n}
+{1\over2}\kappa_{i;n}x\widehat P_{i;n-1}
-{1\over4}\kappa_{i;n}\kappa_{i;n-1}\widehat P_{i;n-2}=0 \eqno(2.58)$$
where, $\kappa_{i;n}$ is defined by equation (2.43).
This relation still holds for $n=i$ and $n=i+1$, since $\kappa_{i;n}$ vanishes for $n=i$.
It has to be compared with the three term recurrence formula (A.16).
For $i=0$, with equation (2.26), one finds
$$H_{n+2}-2\,x\,H_{n+1}+2\,H_n+4n\,x\,H_{n-1}-4n(n-1)\,H_{n-2}=0 \eqno(2.59)$$
which follows from the three term recurrence formula (A.16).

Note that the five term recurrence formula (2.58) can as well be obtained from the first and second order differentiation
formulae (2.44) and (2.56) as follows.
Take the derivative of equation (2.44), eliminate $\widehat P''_{i;n}$ using equation (2.56), and then eliminate
the remaining first derivatives $\widehat P'_{i;n+1}$, $\widehat P'_{i;n}$ and $\widehat P'_{i;n-1}$ using
equation (2.44) again.

As the differentiation formulae (2.44) and (2.50), the recurrence formula (2.58)
can be also used to get step by step $\widehat P_{i;n}$ for all $i$ and $n$ with $0\le i\le n$,
starting from $\widehat P_{i;i}$ (see appendix E).

\bigskip
\noindent {\it 2.5. Recurrence relations with respect to both $i$ and $n$}

\medskip
\noindent (i) As shown in section 3.3.6 of paper I, SBO polynomials corresponding to different
values of $i$ can be related together.
Thus, $\widehat P_{2i+2;2n}$ can be expressed in terms of the polynomials $\widehat P_{2i,2\ell}$
with $\ell=i,\ldots,n$.
From equations I-(3.58), (2.20) and (2.24), one has for $0\le i<n$,
$$\widehat P_{2i+2;2n}
=\sum_{\ell=i}^n \widehat P_{2i;2\ell} {2^{\ell-n}\over n-i} {n-i\choose n-\ell}
{\Gamma\bigl({3\over2}+i+n\bigr)\over\Gamma\bigl({1\over2}+i+\ell\bigr)}
\sum_{m=\sup(i+1,\ell)}^n  (-1)^{m-\ell} {n-\ell\choose m-\ell} {m-i\over{1\over2}+i+m} \eqno(2.60)$$
where, the sum over $m$ can be evaluated as follows.
First, the lowest value of $m$ can be taken as $\ell$ instead of $\sup(i+1,\ell)$, since the term added thereby
occurs for $\ell=i$ only, and it vanishes for $m=i$.
Then, with $p=n-\ell$, $q=m-\ell$, $\rho=\ell-i$ and $\sigma={1\over2}+i+\ell>0$, the sum over $m$ reads
$$\sum_{q=0}^p (-1)^q {p\choose q} {q+\rho\over q+\sigma}
=\delta_{p,0}+(\rho-\sigma) \sum_{q=0}^p (-1)^q {p\choose q} {1\over q+\sigma} \eqno(2.61)$$
where, using the binomial formula and the {\it beta function} [3]
\note{See, e.g., [3] 6.2.1 and 6.2.2.},
$$\int_0^1 dx\,x^{\sigma-1} (1-x)^p=\sum_{q=0}^p (-1)^q {p\choose q} {1\over q+\sigma}
=B(\sigma,p+1)={\Gamma(\sigma) \Gamma(p+1)\over\Gamma(\sigma+1+p)}\,. \eqno(2.62)$$
Putting all together, one finds for $0\le i<n$,
$$\widehat P_{2(i+1);2n}=\widehat P_{2i;2n}-2^{-n} \bigl({\textstyle{1\over2}}+2i\bigr) (n-i-1)!
\,\sum_{\ell=i}^{n-1} {2^\ell\over(\ell-i)!} \widehat P_{2i;2\ell}\,. \eqno(2.63)$$
Similarly for the odd polynomials, from equations (2.21) and (2.25), one has for $0\le i<n$,
$$\widehat P_{2(i+1)+1;2n+1}=\widehat P_{2i+1;2n+1}-2^{-n} \bigl({\textstyle{3\over2}}+2i\bigr) (n-i-1)!
\,\sum_{\ell=i}^{n-1} {2^\ell\over(\ell-i)!} \widehat P_{2i+1;2\ell+1}\,. \eqno(2.64)$$

With always the parity relation (2.4), these recurrence relations provide another way of computing
$\widehat P_{i;n}$ for all $0\le i\le n$, starting with the boundary values $\widehat P_{0;n}$
given by equation (2.26) (see appendix E).
Thus, for $i=0$, one gets for $n\ge1$ [6]
\note{Equation (2.65) has been already given in [6] (3) with different choices of $w$ and $w_2$, see footnote 50.},
$$\eqalignno{
\widehat P_{1;2n}(x)=\widehat P_{2;2n}(x) &=2^{-3n} H_{2n}(\sqrt2 x)-2^{-(n+1)} (n-1)!
\,\sum_{\ell=0}^{n-1} {2^{-2\ell}\over\ell!} H_{2\ell}(\sqrt2 x) &(2.65)\cr
\widehat P_{2;2n+1}(x)=\widehat P_{3;2n+1}(x) &=2^{-(3n+{3\over2})} H_{2n+1}(\sqrt2 x)-2^{-(n+1)} 3(n-1)!
\,\sum_{\ell=0}^{n-1} {2^{-(2\ell+{3\over2})}\over\ell!} H_{2\ell+1}(\sqrt2 x) &(2.66)\cr
}$$
i.e. expansions in terms of Hermite polynomials of the variable $\sqrt2 x$, instead of $x$
as given by equations (2.20) and (2.21).

\smallskip
\noindent (ii) On the right-hand side of equation (2.63), the sum over $\ell$ depends on $n$ through the
upper bound $n-1$ of the sum only.
Therefore, replacing $n$ by $n-1$ in equation (2.63) and eliminating $\sum_{\ell=i}^{n-2} \cdots$
between the equation obtained thereby and equation (2.63) yield for $0\le i\le n-1$,
$$4\widehat P_{2(i+1);2n}-4\widehat P_{2i;2n}-2(n-i-1) \widehat P_{2(i+1);2(n-1)}
+(2i+2n-1) \widehat P_{2i;2(n-1)}=0\,. \eqno(2.67)$$
This relation still holds for $n=i+1$ since then, it reduces to equation (2.63).
From equation (2.64), one gets a similar equation for the odd polynomials.
Actually, both equations can be written as a remarkable single {\sl linear four term recurrence
relation, with coefficients independent of $x$, and which holds for all $i$ and $n$ having the same parity
and such that $0\le i\le n-2$},
$${\rm if}\ (-1)^{i+n}=1\quad{\rm then}\quad
4\widehat P_{i+2;n}-4\widehat P_{i;n}-(n-i-2)\widehat P_{i+2;n-2}+(i+n-1)\widehat P_{i;n-2}=0\,. \eqno(2.68)$$
For $n=i+2$, this relation reduces to a three term recurrence relation.
From equations (2.4) and (2.22), one finds for $i\ge0$, the following expressions which
can also be obtained from equations (2.20) and (2.21),
$$\eqalignno{
\widehat P_{i;i+2} &=2^{-(i+2)}\bigl(H_{i+2}+(1+2i)H_i\bigr) &(2.69)\cr
\noalign{\smallskip}
\widehat P_{i+1;i+3}=\widehat P_{i;i+3} &=2^{-(i+3)}\bigl(H_{i+3}+(3+2i)H_{i+1}\bigr)\,. &(2.70)\cr
}$$

More generally, the four term recurrence relation (2.68) with the parity relation (2.4)
provide another easy way of computing step by step $\widehat P_{i;n}$ for all $i$ and $n$ with $0\le i\le n$,
starting from the boundary values of $\widehat P_{i;i}$ given by equation (2.22) (see appendix E).

\bigskip
\noindent {\it 2.6. Special value at $x=0$}

\medskip
\noindent The odd polynomials $\widehat P_{2i;2n+1}=\widehat P_{2i+1;2n+1}$ vanish at $x=0$.
For the even polynomials, from equations (2.20), (A.11) and (C.1), one finds for $0\le i\le n$,
$$\eqalignno{
\widehat P_{2i-1;2n}(0)=\widehat P_{2i;2n}(0) &=2^{-n} \sum_{m=i}^n (-1)^m 2^m {n-i\choose m-i}
{\Gamma\bigl({1\over2}+i+n\bigr)\over\Gamma\bigl({1\over2}+i+m\bigr)} \bigl({\textstyle{1\over2}}\bigr)_m \cr
\noalign{\smallskip}
&=(-1)^i\,2^{-(n-i)} \bigl(\textstyle{1\over2}\bigr)_i
\bigl(\textstyle{1\over2}+2i\bigr)_{n-i}
{\rm F}\bigl(-(n-i),\textstyle{1\over2};\textstyle{1\over2}+2i;2\bigr) &(2.71)\cr
}$$
where, $\rm F$ is a terminating hypergeometric function, for which we don't know a closed form in terms of
say gamma function products for all $i$ and $n$.
These values can be computed by recurrence as follows.

\smallskip
\noindent {\it (i) Boundary values for $n=i$ and $i=0$.}
From equations (2.22), (2.26) and (A.11), one gets
$$\eqalignno{
\widehat P_{2i;2i}(0) &=2^{-2i} H_{2i}(0)=(-1)^i \bigl({\textstyle{1\over2}}\bigr)_i
=(-1)^i {(2i)!\over2^{2i} i!}\quad i=0,1,\ldots &(2.72)\cr
\widehat P_{0;2n}(0) &=2^{-3n} H_{2n}(0)=(-1)^n 2^{-n} \bigl({\textstyle{1\over2}}\bigr)_n
=(-1)^n {(2n)!\over2^{3n} n!}\quad n=0,1,\ldots\,. &(2.73)\cr
}$$
Setting for $0\le i\le n$,
$$\widehat P_{2i;2n}(0):=(-1)^n 2^{-2(n-i)} \bigl({\textstyle{1\over2}}\bigr)_i\,p_{i;n}
=(-1)^n {(2i)!\over2^{2n} i!} p_{i;n} \eqno(2.74)$$
one finds the boundary values,
$$p_{i;i}=1\quad i=0,1,\ldots
\qquad p_{0;n}=2^n \bigl({\textstyle{1\over2}}\bigr)_n={(2n)!\over2^n n!}\quad n=0,1,\ldots\,. \eqno(2.75)$$
\noindent {\it (ii) Recurrence relation with respect to $n$.}
Setting $x=0$ either in the differentiation equation (2.50), or in the recurrence formula (2.58),
yields for $0\le i\le n$ (setting $\widehat P_{2i;2(i-1)}(0):=0$),
$$4\widehat P_{2i;2(n+1)}(0)+\widehat P_{2i;2n}(0)
-(n-i)\bigl(n+i-\textstyle{1\over2}\bigr)\widehat P_{2i;2(n-1)}(0)=0 \eqno(2.76)$$
or equivalently from equation (2.74), $p_{i;n}$ satisfies the three term recurrence relation
(setting $p_{i;i-1}:=0$),
$$p_{i;n+1}-p_{i;n}-2(n-i)(2n+2i-1)p_{i;n-1}=0\,. \eqno(2.77)$$
With the boundary values (2.75), it follows by recurrence that
{\sl for $m\ge0$, $p_{i;i+m}$ is a polynomial in $i$ of degree $[m/2]$ with positive integer coefficients},
e.g.:
$$p_{i;i+1}=1
\qquad p_{i;i+2}=8\,i+3
\qquad p_{i;i+3}=24\,i+15
\qquad p_{i;i+4}=192\,i^2+336\,i+105\,. \eqno(2.78)$$
\noindent {\it (ii) Recurrence relation with respect to both $i$ and $n$.}
Setting $x=0$ in equation (2.67) yields for $0\le i\le n-1$ (setting $\widehat P_{2(i+1);2i}(0):=0$),
$$4\widehat P_{2(i+1);2n}(0)-4\widehat P_{2i;2n}(0)-2(n-i-1)\widehat P_{2(i+1);2(n-1)}(0)
+(2i+2n-1)\widehat P_{2i;2(n-1)}(0)=0 \eqno(2.79)$$
or equivalently with equation (2.74) (setting $p_{i+1;i}:=0$),
$$2(2i+1)p_{i+1;n}-p_{i;n}+4(2i+1)(n-i-1)p_{i+1;n-1}-(2i+2n-1)p_{i;n-1}=0 \eqno(2.80)$$
allowing also to get step by step $p_{i;n}$ for all $i$ and $n$ such that $0\le i\le n$
starting from the boundary values (2.75).

\smallskip
Note that setting $x=0$ in equation (2.44) (resp. (2.56)) allows to compute $\widehat P'_{i;n}(0)$
(resp. $\widehat P''_{i;n}(0)$), knowing $\widehat P_{i;n}(0)$ for several values of $n$.

\bigskip
\noindent {\it 2.7. Properties of the zeros}

\medskip
\noindent Several properties of the zeros of standard orthogonal polynomials have been known for a long time:
e.g., as recalled in I-(3.3.7),

\smallskip
\noindent {\sl (i) all the zeros are real, simple and located in the support $[a,b]$ of the measure;

\noindent (ii) the zeros of polynomials with consecutive degrees separate each other}.

\smallskip
We saw previously that Hermite SBO polynomials coincide in several special cases with standard
orthogonal polynomials, and therefore share their properties,
i.e. from equations (2.4), (2.22), (2.26) and (2.32),
$$\eqalignno{
\widehat P_{0;0} &=\widehat H_0=1
\qquad\widehat P_{n;n}=\widehat P_{n-1;n}=\widehat H_n \quad n=1,2,\ldots&(2.81)\cr
\noalign{\smallskip}
\widehat P_{0;n}(x) &=2^{-{1\over2}n} \widehat H_n(\sqrt2\,x)
\qquad\widehat P_{1;2n+1}(x)=2^{-3(n+{1\over2})}\widehat H_{2n+1}(\sqrt2 x)\quad n=0,1,\ldots\,. &(2.82)\cr
}$$

\smallskip
As a general result proven in section 3.3.7 of paper I:

\smallskip
\noindent {\sl (i) for  $0\le i\le n$, $\widehat P_{i;n}$ has at least $m$ distinct real zeros of odd order,
with $i\le m\le n$;

\noindent (ii) $\widehat P_{n-1;n},\,n=1,2,\ldots$ has $n$ real and simple zeros}.

\smallskip
It is recalled that the arguments used for standard orthogonal polynomials
can no longer be used since the scalar product $(\,,\,)_0$, defined by equation I-(3.5) (see footnote 12),
is not defined by a single integral.
We observe that all special Hermite SBO polynomials of degree $n$
listed in appendix E have numerically $n$ real simple zeros.
Nevertheless, we have not been able neither to prove this fact for all $n$, nor to build a counter-example.

Concerning the relative positions of the zeros of SBO polynomials with a given $i$ and consecutive
degrees, we only observe numerically that the zeros of the special Hermite SBO polynomials listed in appendix E
separate each other.
Once again, we have not been able to prove a general property.

\medskip
\bigskip
\noindent{\elevenrmb 3. Laguerre standard block orthogonal polynomials}

\medskip
\noindent The study use the same method and follows the same steps as for the Hermite polynomials in section 2.
The (generalized) Laguerre polynomials $L_n^{(\alpha)},\,n=0,1,\ldots$, see appendix C,
correspond to the measure defined by
$$a:=0\quad b:=\infty\qquad w:=e^{-x}\,x^\alpha\quad\alpha>-1\,. \eqno(3.1)$$

\bigskip
\noindent {\it 3.1. Metric tensor components for a particular weight function $w_2$}

\medskip
\noindent {\it (i) Basic quantity $\gamma_{j,k}$.}
With the same interval $[a,b]$, let us choose,
$$w_2=w_2^{(\mu,\alpha)}:=e^{-\mu x}\,x^\alpha\quad\alpha>-1\quad\mu>0\,. \eqno(3.2)$$
This choice is motivated by the application considered in [7] for the special value $\mu=2$.
Our study will be also restricted in (ii) below to this special case, since it allows to push the calculation
to its end analytically.
Then, with the same steps as for equation (2.6), using the equations (B.6) and (B.3), one gets
$$\gamma_{j,k}(\alpha,\mu):=\Bigl(L_j^{(\alpha)}\,,\,L_k^{(\alpha)}\Bigr)_2
={\Gamma(\alpha+1+j)\over j!} {\Gamma(\alpha+1+k)\over k!}\,\mu^{-(\alpha+1)}
S\bigl(j,k,\alpha+1;-\mu^{-1},-\mu^{-1}\bigr) \eqno(3.3)$$
where, the function $S$ is defined in equation (C.8).
Alternately, the result above can be obtained using the definite integral given in [5, 10]
\note{See, [5] 7.414 4. or [10] 2.19.14.6. p 477, using also the linear transformation formula
[9] 15.3.6.}.
Another way to get the basic quantity $\gamma_{j,k}(\alpha,\mu)$ is to compute the scalar product
$(F_s^{(\alpha)}\,,\,F_t^{(\alpha)})_2$ where, $F_z^{(\alpha)}(x)$ is the generating function of
Laguerre polynomials given by equation (B.17).
Then, $\gamma_{j,k}(\alpha,\mu)$ is the coefficient of $s^j t^k$ in the expansion of this scalar product,
i.e.,
$$\Bigl(F_s^{(\alpha)}\,,\,F_t^{(\alpha)}\Bigr)_2
=\sum_{j,k=0}^\infty \gamma_{j,k}(\alpha,\mu) s^j t^k
=\Gamma(\alpha+1) \bigl(\mu+(\mu-2) st-(\mu-1) (s+t)\bigr)^{-\alpha-1}\,. \eqno(3.4)$$

For $\mu=1$, i.e. for $w_2=w$, using equation (C.14), one recovers the orthogonality relation of
Laguerre polynomials,
$\gamma_{j,k}(\alpha,1)=h_j\,\delta_{j,k}$, with $h_j$ given by equation (B.5).
This result follows also from equation (3.4), using equation (C.1),
$$\Bigl(F_s^{(\alpha)}\,,\,F_t^{(\alpha)}\Bigr)_2=\Gamma(\alpha+1) (1-st)^{-\alpha-1}
=\Gamma(\alpha+1) \sum_{j=0}^\infty (-1)^j {-\alpha-j\choose j} (st)^j
=\sum_{j=0}^\infty h_j (st)^j\,. \eqno(3.5)$$

\medskip
\noindent {\it (ii) The special case $w_2=w_2^{(2,\alpha)}=e^{-2x} x^\alpha$.}
As for Hermite polynomials, when $\mu=2$ one gets simple expressions.
From equations (3.3) and (C.15) setting $z=-{1\over2}$, one finds
$$\gamma_{j,k}(\alpha,2)=2^{-(\alpha+1+j+k)} {\Gamma(\alpha+1+j+k)\over j!\,k!} \eqno(3.6)$$
i.e., once again, the same structure with a factor $\Gamma(c+j+k)$ multiplied by some factors depending
separately on $j$ and $k$.
Alternately, the expression above follows also from equation (3.4),
$$\eqalignno{
\Bigl(F_s^{(\alpha)}\,,\,F_t^{(\alpha)}\Bigr)_2 &=\Gamma(\alpha+1) (2-s-t)^{-\alpha-1} \cr
&=\Gamma(\alpha+1) \sum_{n=0}^\infty \sum_{m=0}^n (-1)^n {-\alpha-1\choose n} {n\choose m}
2^{-\alpha-1-n} s^m t^{n-m}\,. &(3.7)
}$$
Then, setting $j=m$, $k=n-m$ and using equation (C.1), one recovers equation (3.6).

\medskip
\noindent {\it From now on, except explicitly noted, only this special case $\mu=2$ will be considered.}

\noindent Then, the corresponding polynomials $P_{i;n},\,0\le i\le n$ are called the
{\it Laguerre SBO polynomials}.
The special case $i=1$ has already been considered in [7] with different choices of weight functions
\note{See [7] section 2 where, to define the subspace ${\cal P}_1^\bot$,
instead of considering the orthogonal polynomials $Q_n$ with respect to $(\,,\,)$, the author computes numerically
the ``seed" polynomials $\widetilde Q_n:=x^n-(x^0\,,\,x^n)/(x^0\,,\,x^0),\,n=1,2,\ldots$,
which readily fulfil the constraint $(x^0\,,\,\widetilde Q_n)=0$.
This can be done easily because there is only one constraint, i.e. $i=1$.}.
(See footnote 52 in appendix F for interrelations with our polynomial $\widehat P_{i;n}$.)

\bigskip
\noindent {\it 3.2. Connection coefficients between block orthogonal and classical polynomials}

\medskip
\noindent As for Hermite polynomials, the relevant quantities are given in terms of determinants
which can be evaluated using lemma D.
One finds from equation I-(3.14), for $0\le i\le n$,
$$Z_{i;n}=2^{-(\alpha+i+n+1)(n-i+1)}
{\prod_{j=i}^n (j-i)!\,\Gamma(\alpha+1+i+j)\over\prod_{j=i}^n (j!)^2}
\qquad Z_{i;i-1}:=1 \eqno(3.8)$$
and therefore, from equations I-(3.15), I-(3.7), I-(3.16) and I-(3.17) with $k_n$ given by equation (B.2) [7]
\note{Equations (3.9) (resp. (3.10)) generalizes the formula given in [7] (7) (resp. (A.1))
for the special case $i=1$ with different
choices of $w$ and $w_2$, see equations (F.16) and footnote 52.},
$$\eqalignno{
\widehat H_{i;n} &=2^{-(\alpha+1+2n)} (n-i)!\,\Gamma(\alpha+1+i+n) &(3.9)\cr
\noalign{\smallskip}
\widehat P_{i;n} &=\sum_{m=i}^n (-1)^m 2^{m-n} m!\,{n-i\choose m-i}
{\Gamma(\alpha+1+i+n)\over\Gamma(\alpha+1+i+m)} L_m^{(\alpha)} &(3.10)\cr
L_n^{(\alpha)} &=\sum_{m=i}^n (-1)^m 2^{m-n} {1\over n!} {n-i\choose m-i}
{\Gamma(\alpha+1+i+n)\over\Gamma(\alpha+1+i+m)} \widehat P_{i;m} &(3.11)\cr
}$$
where, according to equation I-(3.13) and (B.2), for $i\ge0$,
$$\widehat P_{i;i}=\widehat L_i^{(\alpha)}:=(-1)^i\,i!\,L_i^{(\alpha)}\,. \eqno(3.12)$$
The orthogonality and normalization of these polynomials with
respect to the scalar product $(\,,\,)_2$ can be checked computing
$\int_0^\infty dx\,e^{-2x}\,x^\alpha\,\widehat P_{i;n}(x)\,\widehat P_{i;n'}(x)$
with the expansion (3.10) and equations (3.6) and (C.14), thereby recovering equation (3.9).

Equations (3.10) and (3.11) should be compared to equations (B.6) and (B.9), respectively.
More precisely, since $w=w_2^{(1,\alpha)}$, following arguments similar to those given for equation (2.30)
\note{What is new in the Laguerre case with respect to the Hermite one is that a shift of $i$ in the powers of
$x$ can be interpreted as a shift in the parameter $\alpha$.
This is why in the Hermite case, this argumentation can be used with $i=0$ only.},
it follows from equations (B.3) and (3.6) that,
$$\eqalignno{
(x^{i+j}\,,\,x^{i+k})^{(1,\alpha)} &=(x^j\,,\,x^k)^{(1,\alpha+2i)}=\Gamma(\alpha+1+2i+j+k) \cr
\noalign{\smallskip}
&=2^{\alpha+1+2i} \Bigl(2^j\,(i+j)!\,L_{i+j}^{(\alpha)}\,,\,2^k\,(i+k)!\,L_{i+k}^{(\alpha)}\Bigr)^{(2,\alpha)}
 &(3.13)\cr
}$$
where, $(\,,\,)^{(\mu,\alpha)}$ denotes the scalar products defined with the weigh
function $w^{(\mu,\alpha)}$, given by equation (3.2).
Thus, the metric tensor components of the scalar product $(\,,\,)^{(1,\alpha+2i)}$ in the basis
$\{x^n,\,n=0,1,\ldots\}$ and the metric tensor components of the scalar product $(\,,\,)^{(2,\alpha)}$
in the basis $\{2^n (i+n)!\,L_{i+n}^{(\alpha)},\,n=0,1,\ldots\}$
are equal apart from a common multiplicative factor.
Hence, while G-SO with respect to $(\,,\,)^{(1,\alpha+2i)}$ of the basis $\{x^n,\,n=0,1,\ldots\}$
defines the Laguerre polynomial $L_n^{(\alpha+2i)}$,
G-SO with respect to $(\,,\,)^{(2,\alpha)}$ of the basis $\{2^n (i+n)!\,L_{i+n}^{(\alpha)},\,n=0,1,\ldots\}$
defines the polynomial $\widehat P_{i;i+n}$.
In other words, for given $i\ge0$, the linear mapping defined by
$x^n\,\mapsto\,2^n (i+n)!\,L_{i+n}^{(\alpha)},\,n=0,1,\ldots$
is such that $L_n^{(\alpha+2i)}\,\mapsto\,\propto\widehat P_{i;i+n}$.
Therefore, apart from a multiplicative factor $c$, $\widehat P_{i,i+n}$ reads as a linear
combination of
$2^m (i+m)!\,L_{i+m}^{(\alpha)},\,m=0,1,\ldots$ with {\it the same coefficients} as in the expansion of
$L_n^{(\alpha+2i)}$ in terms of the monomials, i.e. from equation (B.6),
$$\widehat P_{i;i+n}=c \sum_{m=0}^n {(-1)^m\over n!} {\Gamma(\alpha+2i+1+n)\over\Gamma(\alpha+2i+1+m)}
{n\choose m} 2^m (i+m)!\,L_{i+m}^{(\alpha)}\,. \eqno(3.14)$$
With $\widehat P_{i;i+n}$ a monic polynomial, computing the coefficient of $x^n$ fixes $c=(-1)^i 2^{-n} n!$.
Changing $i+m$ into $m$ and $i+n$ into $n$, with now $0\le i\le n$, one recovers equation (3.10).

For $i=0$, $\widehat P_{0;n}$
has to be the unique standard monic polynomial $\widehat Q_{2;n}$ corresponding to the weight function $w_2$.
Since, $w_2^{(2,\alpha)}(x)=2^{-\alpha}\,w^{(1,\alpha)}(2x)$, with the normalization (B.2), one gets for $n\ge0$,
$$\widehat P_{0;n}(x)=2^{-n} \widehat L_n^{(\alpha)}(2x)=(-1)^n 2^{-n} n!\,L_n^{(\alpha)}(2x)\,. \eqno(3.15)$$
Thereby, setting $i=0$ in equations (3.10) and (3.11) yields the remarkable linear relations between the two basic
sets of polynomials $\{L_m^{(\alpha)},\,m=0,\ldots,n\}$ and $\{L_m^{(\alpha)}(2x),\,m=0,\ldots,n\}$ for $n\ge0$,
$$\eqalignno{
L_n^{(\alpha)}(2x) &=(-1)^n \sum_{m=0}^n (-1)^m 2^m {(\alpha+1+m)_{n-m}\over(n-m)!} L_m^{(\alpha)}(x) &(3.16)\cr
L_n^{(\alpha)}(x) &=2^{-n} \sum_{m=0}^n {(\alpha+1+m)_{n-m}\over (n-m)!}\,L_m^{(\alpha)}(2x)\,. &(3.17)\cr
}$$
These relations can be checked using the integral already quoted in footnote 18
and the orthogonality relation of Laguerre polynomials with $h_n$ given by equation (B.5), e.g., one has
$$\int_0^\infty dx\,e^{-x}\,x^\alpha\,L_n^{(\alpha)}(2x)\,L_m^{(\alpha)}(x)
=(-1)^{n+m}\,2^m\,{(\alpha+1+m)_{n-m}\over(n-m)!}\,h_m\,. \eqno(3.18)$$

It follows from equations I-(3.10), I-(3.11), (3.10) and (B.7) that,
$$\eqalignno{
\widehat R_{i;n} &=-{1\over2} \bigl(i(\alpha+i)+n(\alpha+n)\bigr) &(3.19)\cr
\widehat S_{i;n} &={1\over8} \Bigl(\,i^3(2\alpha+i)+\bigl(2n^2+2(\alpha-3)n-3(\alpha-1)\bigr)\,i(\alpha+i)
+n(n-1)(\alpha+n)(\alpha+n-1)\Bigr)\,. &(3.20)\cr
}$$

For general properties of SBO polynomials about the {\it projection operators}, the {\it integral
representations} and the {\it normalizations}, see sections 3.3.1, 3.3.2 and 3.3.4 of paper I, respectively.

\bigskip
\noindent {\it 3.3. Differentiation formulae}

\medskip
\noindent For the scalar product $(\,,\,)$, the weight function $w$ satisfies the differential equation (B.11),
which implies the relations (B.12) and (B.13).
Similarly, for the scalar product $(\,,\,)_2$ corresponding to the weight function $w_2^{(\mu,\alpha)}$
defined by equation (3.2), one has for any polynomials $f$ and $g$,
$$\eqalignno{
(x w_2^{(\mu,\alpha)})' &=-(\mu x-\alpha-1) w_2^{(\mu,\alpha)} &(3.21)\cr
\noalign{\smallskip}
(f\,,\,x g')_2 &=-\bigl(x f'-(\mu x-\alpha-1) f\,,\,g\bigr)_2 &(3.22)\cr
\noalign{\smallskip}
\bigl(x f''-(\mu x-\alpha-1) f'\,,\,g\bigr)_2 &=-\bigl(f'\,,\,x\,g'\bigr)_2
=(f\,,\,x g''-(\mu x-\alpha-1) g'\bigr)_2\,. &(3.23)\cr
}$$
\noindent (i) For $0\le\ell<i-1$, the polynomial $x{L_\ell^\alpha}'$ of degree $\ell$
is in the subspace ${\cal P}_i$.
Therefore, $x{L_\ell^\alpha}'$ and $\widehat P_{i;n}$ for $i\le n$ are orthogonal with respect to $(\,,\,)$.
From equation (B.12), one has for $0\le\ell\le i-1$ and $0\le i\le n$,
$$0=\Bigl(\widehat P_{i;n}\,,\,x{L_\ell^{(\alpha)}}'\Bigr)
=-\Bigl(x\widehat P'_{i;n}-(x-\alpha-1)\widehat P_{i;n}\,,\,L_\ell^{(\alpha)}\Bigr)
=-\Bigl(x\widehat P'_{i;n}-x\widehat P_{i;n}\,,\,L_\ell^{(\alpha)}\Bigr)\,. \eqno(3.24)$$
Hence, the particular polynomial $x\widehat P'_{i;n}-x\widehat P_{i;n}$ of degree $n+1$ is orthogonal to
the subspace ${\cal P}_i$ with respect to $(\,,\,)$, and thus reads
$\sum_{j=i}^{n+1} \alpha_{i;n,j} \widehat P_{i;j}$.
Computing the coefficient of $x^{n+1}$ in $x\widehat P'_{i;n}-x\widehat P_{i;n}$ fixes $\alpha_{i;n,n+1}=-1$.
Now, with respect to the scalar product $(\,,\,)_2$, one has from equations (1.2) and (3.22), for $i\le j\le n+1$,
$$\eqalignno{
\alpha_{i;n,j} \widehat H_{i;j} &=\bigl(\widehat P_{i;j}\,,\,x\widehat P'_{i;n}-x\widehat P_{i;n}\bigr)_2
=\bigl(\widehat P_{i;j}\,,\,x\widehat P'_{i;n}\bigr)_2-\bigl(x\widehat P_{i;j}\,,\,\widehat P_{i;n}\bigr)_2 \cr
\noalign{\smallskip}
&=-\bigl(x\widehat P'_{i;j}-(\mu-1)x\widehat P_{i;j}\,,\,\widehat P_{i;n}\bigr)_2
-(\alpha+1) \widehat H_{i;n} \delta_{j,n}\,. &(3.25)\cr
}$$
In the special case $\mu=2$, it turns out that $x\widehat P'_{i;j}-(\mu-1)x\widehat P_{i;j}$
is the particular polynomial $x\widehat P'_{i;j}-x\widehat P_{i;j}$, introduced above for $j=n$
(i.e. of degree $j+1$ in the subspace ${\cal P}_i^\bot$), and which reads
$\sum_{k=i}^{j+1} \alpha_{i;j,k} \widehat P_{i;k}$.
Thus, for $\mu=2$, {\it considered henceforth}, one has
$$\alpha_{i;n,j} \widehat H_{i;j}
=-\Bigl(\alpha_{i;j,n}\,\sum_{k=i}^{j+1} \delta_{k,n}+(\alpha+1) \delta_{j,n}\Bigr) \widehat H_{i;n}\,.
\eqno(3.26)$$
It follows that for $j=i,\ldots,n-2$, $\alpha_{i;n,j}$ vanishes.
It remains to consider the cases $j=n-1$ and $j=n$.
With $\alpha_{i;n-1,n}=-1$ and equation (3.9), one gets
$$\kappa_{i;n}(\alpha):=\alpha_{i;n,n-1}={\widehat H_{i;n}\over\widehat H_{i;n-1}}={1\over4}(n-i)(\alpha+i+n)
\qquad\alpha_{i;n,n}=-{1\over2}(\alpha+1)\,. \eqno(3.27)$$
Finally, for $\mu=2$ and $0\le i\le n$, one finds the {\it first order differentiation formula} [7]
\note{This generalizes the formula given in [7] (3) for the special case $i=1$ with different choices of
$w$ and $w_2$, see footnote 52.},
$$x\widehat P'_{i;n}=-\widehat P_{i;n+1}+\,\Bigl(x-{1\over2}(\alpha+1)\Bigr) \widehat P_{i;n}
+\kappa_{i;n}(\alpha) \widehat P_{i;n-1} \eqno(3.28)$$
which still holds for $n=i$, since then $\kappa_{i;n}(\alpha)$ vanishes.
This equation has to be compared with the differentiation formula (B.16).
For $i=0$, using equation (3.15), one gets
$$2 x{L_n^{(\alpha)}}'
=(n+1) L_{n+1}^{(\alpha)}+(x-\alpha-1) L_n^{(\alpha)}-(n+\alpha) L_{n-1}^{(\alpha)} \eqno(3.29)$$
which follows from, e.g., adding the two differentiation formulae (B.14) and (B.16).

With the boundary value of $\widehat P_{i;i}$ given in terms of Laguerre polynomials by equation (3.12),
the differentiation formula (3.28) allows to get step by step $\widehat P_{i;n}$ for all $i$ and $n$
with $0\le i\le n$.
A table of these polynomials for $i=0,1,2$ and $n=i,i+1,\ldots,i+4$ is given in appendix F.
In particular, setting $n=i+1$ and using equations (3.12), (B.2) and (B.16), one finds for $n=i\ge0$,
$$\widehat P_{i;i+1}=\widehat L_{i+1}^{(\alpha)}+{1\over2}(\alpha+1+2i)\,\widehat L_i^{(\alpha)}
=(-1)^{i+1}(i+1)!\,\,L_{i+1}^{(\alpha)}+(-1)^i\,{1\over2}\,i!(\alpha+1+2i)\,L_i^{(\alpha)}\,. \eqno(3.30)$$
This result can be obtained directly setting $n=i+1$ in equation (3.10).

\noindent (ii) Following the same method again,
for $0\le\ell<i\le n$, the polynomial $x{L_\ell^\alpha}''-(x-\alpha-1){L_\ell^\alpha}'$ of degree $\ell$
is orthogonal to $\widehat P_{i;n}$ with respect to the scalar product $(\,,\,)$,
and from equation (B.13), one has
$$0=\Bigl(\widehat P_{i;n}\,,\,x{L_\ell^{(\alpha)}}''-(x-\alpha-1){L_\ell^{(\alpha)}}'\Bigr)
=\Bigl(x\widehat P''_{i;n}-(x-\alpha-1)\widehat P'_{i;n}\,,\,L_\ell^{(\alpha)}\Bigr)\,. \eqno(3.31)$$
Hence, the particular polynomial $x\widehat P''_{i;n}-(x-\alpha-1)\widehat P'_{i;n}$ of degree $n$ belongs to
${\cal P}_i^\bot$  and reads $\sum_{j=i}^n\gamma_{i;n,j} \widehat P_{i;j}$.
Computing the coefficient of $x^{n}$ in $x\widehat P''_{i;n}-(x-\alpha-1)\widehat P'_{i;n}$ fixes
$\gamma_{i;n,n}=-n$.
With the scalar product $(\,,\,)_2$, one has from equations (1.2), (3.22) and (3.23), for $i\le j\le n$,
$$\eqalignno{
\gamma_{i;n,j} \widehat H_{i;j}
&=\bigl(\widehat P_{i;j}\,,\,x\widehat P''_{i;n}-(x-\alpha-1)\widehat P'_{i;n}\bigr)_2 \cr
\noalign{\smallskip}
&=\bigl(\widehat P_{i;j}\,,\,x\widehat P''_{i;n}-(\mu x-\alpha-1)\widehat P'_{i;n}\bigr)_2
+(\mu-1) \bigl(\widehat P_{i;j}\,,\,x\widehat P'_{i;n}\bigr)_2 \cr
\noalign{\smallskip}
&=\Bigl(x\widehat P''_{i;j}-\bigl((\mu-1) x-\alpha-1\bigr)\widehat P'_{i;j}\bigr)\,,\,\widehat P_{i;n}\Bigr)_2
-\mu \Bigl(x\widehat P'_{i;j}-(\mu-1) x\widehat P_{i;j}\bigr)\,,\,\widehat P_{i;n}\Bigr)_2 \cr
&\phantom{=}-(\mu-1)(\alpha+1) \widehat H_{i;n} \delta_{j,n}\,. &(3.32)\cr
}$$
In the special case $\mu=2$, once again, $x\widehat P''_{i;j}-\bigl((\mu-1) x-\alpha-1\bigr)\widehat P'_{i;j}$
(resp. $x\widehat P'_{i;j}-(\mu-1) x\widehat P_{i;j}$) is the particular polynomial
$x\widehat P''_{i;j}-(x-\alpha-1)\widehat P'_{i;j}$ (resp. $x\widehat P'_{i;j}-x\widehat P_{i;j}$),
introduced above for $j=n$
(i.e. of degree $j$ (resp. $j+1$) in the subspace ${\cal P}_i^\bot$), and which reads
$\sum_{k=i}^j \gamma_{i;j,k} \widehat P_{i;k}$ (resp. $\sum_{k=i}^{j+1} \alpha_{i;j,k} \widehat P_{i;k})$.
Thus, one has
$$\gamma_{i;n,j} \widehat H_{i;j}
=\Bigl(\gamma_{i;j,n} \sum_{k=i}^j \delta_{k,n}-2 \alpha_{i;j,n} \sum_{k=i}^{j+1} \delta_{k,n}
-(\alpha+1) \delta_{j,n}\Bigr) \widehat H_{i;n}\,. \eqno(3.33)$$
For $n=i$, and therefore $j=i$, it is already known that $\gamma_{i;i,i}=-i$.
For $n\ge i+1$, it follows from above that for $n\ge i+2$ and $j=i,\ldots,n-2$, $\gamma_{i;n,j}$ vanishes.
It remains to consider the case $j=n-1$.
From $\alpha_{i;n-1,n}=-1$ and  equation (3.27), one gets
$$\gamma_{i;n,n-1}=-2  \alpha_{i;n-1,n} {\widehat H_{i;n}\over\widehat H_{i;n-1}}
=2 \kappa_{i;n}(\alpha)\,. \eqno(3.34)$$
Finally, in the case $\mu=2$, one finds the {\it second order differentiation formula} for $0\le i\le n$ [7]
\note{This generalizes the formula given in [7] (6) for the special case $i=1$ with different choices of
$w$ and $w_2$, see footnote 52.},
$$x\widehat P''_{i;n}+(\alpha+1-x) \widehat P'_{i;n}+n \widehat P_{i;n}
=2 \kappa_{i;n}(\alpha) \widehat P_{i;n-1} \eqno(3.35)$$
where, $\kappa_{i;n}(\alpha)$ is defined by equation (3.27).
For $n=i$, the relation still holds, since then $\kappa_{i;n}(\alpha)$ vanishes,
and using equation (3.12), the result coincides with the differential equation (B.15).
For $i=0$, using equation (3.15), the relation follows, e.g., from equations (B.14) and (B.15).

\bigskip
\noindent {\it 3.4. Recurrence relation with respect to the degree $n$}

\medskip
\noindent Taking the derivative of the first order differentiation formula (3.28), and eliminating
$x \widehat P''_{i;n}$ with the second order differentiation formula (3.35) yield
$$-\widehat P'_{i;n+1}+{1\over2}(\alpha-1)\widehat P'_{i;n}+(n+1)\widehat P'_{i;n}
+{1\over4}(n-i)(\alpha+1+i+n)(\widehat P'_{i;n-1}-2\widehat P_{i;n-1})=0\,. \eqno(3.36)$$
Then, multiplying both sides of this equation by $x$ and using again equation (3.28) to eliminate the first order
derivatives $\widehat P'_{i;n+1}$, $\widehat P'_{i;n}$
\note{Keeping the term $\widehat P'_{i;n}$ provides another first order differentiation formula.}
and $\widehat P'_{i;n-1}$ give the {\it five term linear recurrence formula} for $0\le i\le n$,
$$\eqalignno{
&\widehat P_{i;n+2}-(x-1)\widehat P_{i;n+1}
+\Bigl({1\over2}(\alpha+1+2n)x-2\kappa_{i;n}(\alpha)-{1\over4}(\alpha^2+\alpha+2n)\Bigr)\widehat P_{i;n} \cr
&-\kappa_{i;n}(\alpha)(x+1)\widehat P_{i;n-1}
-\kappa_{i;n}(\alpha)\kappa_{i;n-1}(\alpha)\widehat P_{i;n-2}=0 &(3.37)\cr
}$$
where, $\kappa_{i;n}(\alpha)$ is defined by equation (3.27).
A direct calculation shows that this relation still holds for $n=i+1$ and $n=i$, where it reduces to a four and
three term recurrence formula, respectively.
For $n=i$, using equation (3.30), one finds for $i\ge0$,
$$\widehat P_{i;i+2}=(x-1)\widehat L_{i+1}^{(\alpha)}
+{1\over4}(\alpha^2-\alpha-2-2i)\widehat L_i^{(\alpha)}\,. \eqno(3.38)$$

As the differentiation formula (3.28), the recurrence formula (3.37) can be used to get step by step
$\widehat P_{i;n}$ for all $i$ and $n$ with $0\le i\le n$ starting from $\widehat P_{i;i}$ and
$\widehat P_{i;i+1}$ given by equations (3.12) and (3.30), respectively (see appendix F).

\bigskip
\noindent {\it 3.5. Recurrence relations with respect to both $i$ and $n$}

\medskip
\noindent (i) Setting $j=i+1$ in equation I-(3.58) and using equations (3.10) and (3.11),
one can express $\widehat P_{i+1;n}$ in terms of the polynomials $\widehat P_{i;\ell}$ with $i\le\ell\le n$.
The calculation is very similar to what was done for Hermite polynomials in section 2.5.
One finds
$$\widehat P_{i+1;n}
=\sum_{\ell=i}^n \widehat P_{i,\ell} {2^{\ell-n}\over n-i} {n-i\choose n-\ell}
{\Gamma(\alpha+2+i+n)\over\Gamma(\alpha+1+i+\ell)}
\sum_{m=\sup(i+1,\ell)}^n  (-1)^{m-\ell} {n-\ell\choose m-\ell} {m-i\over\alpha+1+i+m} \eqno(3.39)$$
and using equations (2.61) and (2.62)
(with $p=n-\ell$, $\rho=\ell-i$ and $\sigma=\alpha+1+i+\ell>0$)
to compute the sum over $m$ yield for $0\le i< n$,
$$\widehat P_{i+1;n}=\widehat P_{i;n}-2^{-n}(\alpha+1+2i)(n-i-1)!
\,\sum_{\ell=i}^{n-1} {2^\ell\over(\ell-i)!} \widehat P_{i;\ell}\,. \eqno(3.40)$$

This recurrence relation provide another way of computing step by step $\widehat P_{i;n}$ for all $i$ and $n$ with
$0\le i\le n$, starting with the boundary values $\widehat P_{0;n}$
given by equation (3.15) (see appendix F).
Thus, for $i=0$, one gets for $n\ge1$,
$$\widehat P_{1;n}=(-1)^n 2^{-n} L_n^{((\alpha)}(2x)-2^{-n} (\alpha+1) (n-1)!
\,\sum_{\ell=0}^{n-1} (-1)^\ell L_\ell^{(\alpha)}(2x) \eqno(3.41)$$
i.e. an expansion in terms of Laguerre polynomials of the variable $2x$, instead of $x$
as given by equation (3.10).

\noindent (ii) On the right-hand side of equation (3.40), the sum over $\ell$ depends on $n$ through the
upper bound $n-1$ of the sum only.
Therefore, replacing $n$ by $n-1$ in this equation and eliminating $\sum_{\ell=i}^{n-2} \cdots$
between the equation obtained thereby and equation (3.40) yield for $0\le i\le n-1$,
$$2\widehat P_{i+1;n}-2\widehat P_{i;n}-(n-i-1)\widehat P_{i+1;n-1}
+(\alpha+i+n)\widehat P_{i;n-1}=0\,. \eqno(3.42)$$
This relation still holds for $n=i+1$, where it reduces to a three term recurrence recovering equation (3.40).

This {\it four term linear recurrence relation} provides an easy way of computing step by step $\widehat P_{i;n}$
for all $i$ and $n$ with $0\le i\le n$, starting from the boundary values of $\widehat P_{i;i}$
given by equation (3.12) (see appendix F).

\bigskip
\noindent {\it 3.6. Special value at $x=0$}

\medskip
\noindent From equations (3.10), (B.8) and (C.1), one finds for $0\le i\le n$,
$$\widehat P_{i;n}(0)=(-1)^i 2^{-(n-i)} (\alpha+1)_i (\alpha+1+2i)_{n-i}
{\rm F}(-(n-i),\alpha+1+i;\alpha+1+2i;2) \eqno(3.43)$$
where, $\rm F$ is a terminating hypergeometric function, for which we don't know an explicit expression
for all $i$ and $n$.
These values can be computed by recurrence as follows.

\smallskip
\noindent {\it (i) Boundary values for $n=i$ and $i=0$.}
From equations (3.12), (3.15) and (B.8), one gets
$$\eqalignno{
\widehat P_{i;i}(0) &=(-1)^i i!\,L_i^{(\alpha)}(0)=(-1)^i (\alpha+1)_i\quad i=0,1,\ldots &(3.44)\cr
\noalign{\smallskip}
\widehat P_{0;n}(0) &=(-1)^n 2^n n!\,L_n^{(\alpha)}(0)=(-1)^n 2^n (\alpha+1)_n\quad n=0,1,\ldots\,. &(3.45)\cr
}$$
Setting for $0\le i\le n$
\note{See equation (3.56) below, $p_{i;n}$ defined for Hermite polynomials by equation (2.74)
and $p_{i;n}^{(\alpha)}$ are related by $p_{i;n}=2^{n-i}\,p_{i;n}^{(-{1\over2})}$.},
$$\widehat P_{i;n}(0):=(-1)^n\,2^{-(n-i)}\,(\alpha+1)_i\,p_{i;n}^{(\alpha)} \eqno(3.46)$$
one finds the boundary values,
$$p_{i;i}^{(\alpha)}=1\quad i=0,1,\ldots
\qquad p_{0;n}^{(\alpha)}=(\alpha+1)_n\quad n=0,1,\ldots\,. \eqno(3.47)$$
\noindent {\it (ii) Recurrence relation with respect to $n$.}
Setting $x=0$ in the differentiation equation (3.28) yields for $0\le i\le n$ (setting $\widehat P_{i;i-1}(0):=0$),
$$4\widehat P_{i;n+1}(0)+2(\alpha+1)\widehat P_{i;n}(0)
-(n-i)(\alpha+i+n)\widehat P_{i;n-1}(0)=0 \eqno(3.48)$$
or equivalently from equation (3.46), $p_{i;n}^{(\alpha)}$ satisfies the three term recurrence relation
(setting $p_{i;i-1}^{(\alpha)}:=0$),
$$p_{i;n+1}^{(\alpha)}-(\alpha+1)p_{i;n}^{(\alpha)}-(n-i)(\alpha+i+n)p_{i;n-1}^{(\alpha)}=0\,. \eqno(3.49)$$
With the boundary values (3.47), it follows by recurrence that
{\sl for $m\ge 0$, $p_{i;i+m}^{(\alpha)}$ is a monic polynomial in $\alpha$
of degree $m$ with positive integer coefficients, and also a polynomial in $i$ of degree $[m/2]$}, e.g.:
$$\eqalignno{
p_{i;i+1}^{(\alpha)} &=\alpha+1
\qquad p_{i;i+2}^{(\alpha)}=\alpha^2+3\,\alpha+2(1+i)
\qquad p_{i;i+3}^{(\alpha)}=(\alpha+1)\,\bigl(\alpha^2+5\,\alpha+6(1+i)\bigr) \cr
\noalign{\smallskip}
p_{i;i+4}^{(\alpha)} &=\alpha^4+10\,\alpha^3+(35+12\,i)\,\alpha^2+2(25+18\,i)\,\alpha+12(1+i)(2+i)\,. &(3.50)\cr
}$$

\noindent {\it (iii) Recurrence relation with respect to both $i$ and $n$.}
Setting $x=0$ in equation (3.42) yields for $0\le i\le n-1$ (setting $\widehat P_{i+1;i}(0):=0$),
$$2\widehat P_{i+1;n}(0)-2\widehat P_{i;n}(0)-(n-i-1)\widehat P_{i+1;n-1}(0)
+(\alpha+i+n)\widehat P_{i;n-1}(0)=0 \eqno(3.51)$$
or equivalently with equation (3.46),
$$2(2\alpha+1+i)p_{i+1;n}^{(\alpha)}-p_{i;n}^{(\alpha)}+2(\alpha+1+i)(n-i-1)p_{i+1;n-1}^{(\alpha)}
-(\alpha+i+n)p_{i;n-1}^{(\alpha)}=0 \eqno(3.52)$$
allowing also to get step by step $p_{i;n}^{(\alpha)}$ for all $i$ and $n$ such that $0\le i\le n$
starting from the boundary values (3.47).

\smallskip
Note that setting $x=0$ in equation (3.35) (resp. its derivative with respect to $x$) allows to compute
$\widehat P'_{i;n}(0)$ (resp. $\widehat P''_{i;n}(0)$), knowing $\widehat P_{i;n}(0)$ for several values of $n$.

\bigskip
\noindent {\it 3.7. Properties of the zeros}

\medskip
\noindent From equations (3.12) and (3.15), the particular Laguerre SBO polynomials,
$$\widehat P_{n;n}=\widehat L_n^{(\alpha)}
\qquad\widehat P_{0;n}(x)=2^{-n}\,\widehat L_n^{(\alpha)}(2x)\quad n=0,1,\ldots \eqno(3.53)$$
share the properties, especially of the zeros, of the standard orthogonal polynomials.

It has been shown in section 3.3.7 of paper I that:

\noindent {\sl - for  $0\le i\le n$, $P_{i;n}$ has at least $m$ distinct non-negative real zeros of odd order,
with $i\le m\le n$};

\noindent {\sl - for $n\ge1$, $P_{n-1;n}$ has $n$ non-negative real and simple zeros}.

\noindent We observe that all the special Laguerre SBO polynomials of degree $n$ listed in appendix F
have numerically $n$ real non-negative simple zeros.
Nevertheless, we have not been able neither to prove this fact in general, nor to build a counter-example.

Concerning the relative positions of the zeros of SBO polynomials with a given $i$ and consecutive
degrees, in the Laguerre case one readily finds a counter-example, e.g., with $\alpha=0$,
from equations (F.6) and (F.7), the zeros read
$$\eqalignno{
\widehat P_{1;2} &=x^2-{5\over2}\,x+{1\over2}
\qquad x_1={5-\sqrt{17}\over4}\approx0.219\quad x_2={5+\sqrt{17}\over4}\approx2.281 &(3.54)\cr
\widehat P_{1;3} &=x^3-5\,x^2+5\,x-1
\qquad y_1=2-\sqrt3\approx0.268\qquad y_2=1\quad y_3=2+\sqrt3\approx3.732 &(3.55)\cr
}$$
where, $x_1<y_1<y_2<x_2<y_3$, and therefore the zeros do not separate each other, as for standard orthogonal
polynomials.
Once again, we have not been able to prove a general property.

\bigskip
\noindent {\it 3.8. Interrelations between Hermite and Laguerre cases}

\medskip
\noindent The interrelations between Hermite and Laguerre polynomials are recalled in appendix B (item 6).
Using the relation (B.19) (resp. (B.20)), the expansion (2.20) (resp. (2.21)) for the even (resp. odd)
polynomials in the Hermite case,
and the expansion (3.10) for the Laguerre case, setting $\alpha=-{1\over2}$ (resp. ${1\over2}$),
one finds for $0\le i\le n$,
$$\eqalignno{
\widehat P_{2i-1;2n}^{({\rm Hermite})}(x)=\widehat P_{2i;2n}^{({\rm Hermite})}(x)
&=\widehat P_{i;n}^{({\rm Laguerre}\ \alpha=-{1\over2})}(x^2) &(3.56)\cr
\noalign{\smallskip}
\widehat P_{2i;2n+1}^{({\rm Hermite})}(x)=\widehat P_{2i+1;2n+1}^{({\rm Hermite})}(x)
&=x\,\widehat P_{i;n}^{({\rm Laguerre}\ \alpha={1\over2})}(x^2) &(3.57)\cr
}$$
i.e. the same interrelations as between Hermite and Laguerre polynomials.
Let us recall that, while the choice of $i$ determines the subspace ${\cal P}_i$ of polynomials of degree
at most $i-1$, the complementary and orthogonal subspace ${\cal P}_i^\bot$, equation (1.3),
depends on the choice of the scalar product $(\,,\,)$.
Therefore,  ${\cal P}_i^{({\rm Hermite})\bot}$,
${\cal P}_i^{({\rm Laguerre}\ \alpha=-{1\over2})\bot}$
and ${\cal P}_i^{({\rm Laguerre}\ \alpha={1\over2})\bot}$ are, a priori, different subspaces.

Relations (3.56) and (3.57) allow a cross check of many similar formulae obtained in sections 2 and 3, and also
of the explicit expressions of the Hermite and the Laguerre SBO polynomials listed in appendices E and F,
respectively.

\medskip
\bigskip
\noindent{\elevenrmb Conclusion}

\bigskip
\noindent In a first paper [1] (paper I), we have defined general {\it block orthogonal polynomials}.
Their properties have been investigated, pointing out what can be generalized and ascribed to Euclidean
vector space, endowed with two distinct scalar products, and to polynomial algebra.
A general study of the special case of {\it standard block orthogonal} (SBO) {\it polynomials}, has been made,
emphasizing the similarities to and the differences from the study of the {\it standard orthogonal polynomials}.
For a given $i=0,1,\ldots$, these real polynomials of degree $n$, $P_{i;n},\,n=i,i+1,\ldots$,
are orthogonal to the $i$-dimensional subspace ${\cal P}_i$
of polynomials of degree less than $i$ with respect to a first Euclidean scalar product $(\,,\,)$.
In addition, they are mutually orthogonal with respect to a second Euclidean scalar product $(\,,\,)_2$.
They form an orthogonal basis with respect to $(\,,\,)_2$ of the subspace ${\cal P}_i^\bot$
of codimension $i$, orthogonal to ${\cal P}_i$ with respect to $(\,,\,)$, and complementary to ${\cal P}_i$
with respect to the space of polynomials of any degree.
Based on the Gram-Schmidt orthogonalization (G-SO), we have determined these polynomials uniquely,
apart from a constant multiplicative factor for each polynomial, in terms of determinants which
entries are the metric tensor components of both scalar products considered.

In this second paper, we apply the general results obtained in paper I, assuming
the first scalar product $(\,,\,)$ corresponds to the Hermite (resp. Laguerre) {\it classical polynomials}.
In both cases, it turn out that for a particular choice of the second scalar product $(\,,\,)_2$,
we get explicitly the connection coefficients between the Hermite (resp. Laguerre) polynomials, and what we call
the {\it Hermite} (resp. {\it Laguerre}) {\it SBO polynomials}.
These polynomials satisfy all the general properties derived in section 3.3 of paper I for the projection
operators, the integral representations and the normalizations.
Furthermore, we establish recurrence relations with respect to $n$ and $i$, and differentiation formulae,
allowing a step by step determination of these SBO polynomials.
As for the classical orthogonal polynomials, these relations are based on differential equations satisfied by
the weight functions $w(x)$ and $w_2(x)$ defining, respectively, the two scalar products.
Partial properties for the zeros of these polynomials are also obtained.
We point out the differences with the classical orthogonal polynomials, basically due to the
fact that two, instead of one, distinct scalar products are involved.
As the Hermite and the Laguerre polynomials, the Hermite and the Laguerre SBO polynomials can be related together.
In the Hermite case, the measures considered being symmetric with respect to the origin,
additional parity properties occur.

This paper recovers and improves results already obtained by Giraud {\it et al} [6--8] for $i=1$,
and partly for $i=2$.
In particular, we get new recurrence and differentiation formulae.
We generalize these results making a systematic study for all non-negative integer values of $i$.
As recalled in paper I, $P_{i;n},\,n=i,i+1,\ldots$ provide a basis of the subspace ${\cal P}_i^\bot$,
suitable for applications requiring to take into account $i$ linearly
independent constraints which can be recast into an orthogonality relation with the subspace ${\cal P}_i$
with respect to $(\,,\,)$.
(We have shown in paper I that an extension of similar methods to constraints associated with different scalar
products is not straightforward, and may have no solution, a unique solution or an infinite number of solutions.)
These constrained polynomials have been first considered in [6--8] with $i=1$ to implement the matter conservation
in view of application of the Hohenberg-Kohn functional in some problems of nuclear physics.

Open questions still remain.
Why can we perform analytical calculations only for very special choices of the
weight function $w_2$ (precisely those made in the applications above), once $w$ is chosen as the classical
Hermite (resp. Laguerre) weight function?
Is there any `integrability' structure to discover?
Nevertheless, let us recall that for any proper weight functions, determination of the SBO polynomials is a
well-defined problem.
G-SO is well-suited for numerical calculations and it is not necessary for applications to have analytical
expressions.
We have not been able to get closed formulae for the connection coefficients between $P_{i;n}$ and $x^m$
(we observe in all the special values of $P_{i;n}$ given in appendices E and F,
alternate signs in the expansion in powers of $x$).
Can we establish that, both in Hermite and Laguerre cases, $P_{i;n}$ has $n$ distinct real zeros?
Similarly, do the zeros of the Hermite SBO polynomials with a given $i$ and consecutive degrees separate
each other?
(We establish this is not true for the Laguerre case.)

As a byproduct, for completeness and to underlined the similarities to and the differences from the classical
polynomials, we recover in appendices well-known properties of Hermite and Laguerre polynomials
in a unified way (including the expansions of monomials in terms of classical polynomials,
not given in the standard textbooks).
In addition, also reported in appendix, we have been led to consider several terminating generalized
hypergeometric series with one or more variables often set to unity.

The Gegenbauer and Jacobi polynomials will be similarly considered in a forthcoming paper.

We have already underlined in the conclusion of paper I (see footnote 44 of paper I),
that the general formulae established therein can be used to defined block orthogonal polynomials
for any given $i$-dimensional subspace ${\cal E}_i$, possibly different from ${\cal P}_i$ (see footnote 51).
Then the block orthogonal polynomials would not be the standard ones and require a new similar study.

\medskip
\bigskip
\noindent{\elevenrmb Acknowledgments}

\bigskip
\noindent B Giraud drew my attention to constrained orthogonal polynomials.
I had helpful and friendly discussions about polynomials with M L Mehta at the beginning of this work.
It is a pleasure to thank both of them.

\bigskip
\bigskip
\noindent{\elevenrmb Appendix A. Hermite polynomials}

\bigskip
\noindent It is shown in appendix I-A that all the relevant quantities ($Z_n, h_n, Q_n, a_{m,n},b_{m,n},\ldots$)
can be expressed in terms of determinants with the metric tensor components as elements, except possibly for the last raw,
$$g_{j,k}:=(x^j\,,\,x^k)=c_{j+k}:=\int_a^b dx\,w(x)\,x^{j+k}\,. \eqno(A.1)$$
These general results are now applied to the {\it classical orthogonal polynomials}, i.e. for particular choices
of the measure defining the scalar product.
Thereby, well-known results are recovered using a unique method based on the calculation of some special
determinants with gamma functions as elements (see lemma D in appendix D)
and also the evaluation of some finite sums involving binomial coefficients and ratios of gamma functions
(see lemmas C.1 and C.2 in appendix C).
In addition, since the weight functions $w$ considered satisfy some differential equations, integrations by part
with vanishing boundary conditions yield some relations between scalar products
({\it this is still valid for particular SBO polynomial}, see sections 2.3 and 3.3.).

\medskip
\noindent (1) {\it Definition.}
{\it Hermite polynomials} $H_n,\,n=0,1,\ldots$, are defined with the measure and normalization
\note{See, e.g., [4] 10.13 (1) and (4).},
$$\eqalignno{
a &:=-\infty\qquad b:=\infty\qquad w:={\rm e}^{-x^2} &(A.2)\cr
\noalign{\smallskip}
k_n &:=2^n\quad n=0,1,\ldots\,. &(A.3)\cr
}$$
The measure being symmetric with respect to the origin, $H_n(-x)=(-1)^n H_n(x)$.

\medskip
\noindent (2) {\it Connection coefficients.} From equation (A.1) and the {\it Euler integral}
\note{See, e.g., [3] 6.1.1.},
the nonzero moments read
$$c_{2j}=\int_{-\infty}^\infty dx\,e^{-x^2} x^{2j}
=\Gamma\bigl({\textstyle{1\over2}}+j\bigr)\quad j=0,1,\ldots\,. \eqno(A.4)$$
Therefore, all the determinants which occur in equations I-(A.17)--I-(A.28) can be obtained with lemma D
and the duplication formula (see footnote 6).
For $n=0,1,\ldots$ and $m=0,\ldots,n$, one finds
\note{See, e.g., [4] 10.13 (4) for $h_n$, (9) for the expansion (A.9), (4) for $r_n$ only
(see footnote 47 in paper I), (15) for $H_n(0)$ and (5) for the three term recurrence formula.},
$$\eqalignno{
Z_n^{({\rm e})} &=\prod_{j=0}^n j!\,\Gamma\bigl({\textstyle{1\over2}}+j\bigr)
\qquad Z_n^{({\rm o})}=\prod_{j=0}^n j!\,\Gamma\bigl({\textstyle{3\over2}}+j\bigr) &(A.5)\cr
h_n &=2^n n!\,\pi^{1\over2} &(A.6)\cr
a_{2m,2n} &=k_{2n} (-1)^{n-m}
\,{\Gamma\bigl({\textstyle{1\over2}}+n\bigr)\over\Gamma\bigl({\textstyle{1\over2}}+m\bigr)} {n\choose m}
=(-1)^{n-m} 2^{2m} {(2n)!\over(2m)!\,(n-m)!} &(A.7)\cr
a_{2m+1,2n+1} &=k_{2n+1} (-1)^{n-m}
\,{\Gamma\bigl({\textstyle{3\over2}}+n\bigr)\over\Gamma\bigl({\textstyle{3\over2}}+m\bigr)} {n\choose m}
=(-1)^{n-m} 2^{2m+1} {(2n+1)!\over(2m+1)!\,(n-m)!} &(A.8)\cr
H_n &=\sum_{m=0}^{[n/2]}(-1)^m {n!\over m!\,(n-2m)!} (2x)^{n-2m} &(A.9)\cr
\widehat r_n &=0\qquad\widehat s_n=-{1\over4} n(n-1) &(A.10)\cr
H_{2n}(0) &=(-1)^n 2^{2n} \bigl(\textstyle{1\over2}\bigr)_n=(-1)^n {\displaystyle{(2n)!\over n!}}
\qquad H_{2n+1}(0)=0\,. &(A.11)\cr
}$$
Similarly, the connection coefficient $b_{m,n}$, not given in the standard textbook already quoted,
can be obtained from equations I-(A.10), I-(A.22), I-(A.28), (D.2), (C.3) and
the duplication formula (see footnote 6).
For $m=0,\ldots,n$ one has
\note{See, [3] table 22.12 where, the numerical values of $b_{m,n}$, and also $a_{m,n}$,
are listed up to $n=12$, but the general expressions (A.12)--(A.14) are not given there.
Formulae (A.12) and (A.13) can be found in [11] chapter 6, exercise 6 p 339.},
$$\eqalignno{
b_{2m,2n} &={k_{2m}^{-1}\over m!} \sum_{\ell=0}^m (-1)^{m-\ell} {m\choose\ell}
{\Gamma\bigl({\textstyle{1\over2}}+n+\ell\bigr)\over\Gamma\bigl({\textstyle{1\over2}}+\ell\bigr)} \cr
&=2^{-2m} {\Gamma\bigl({\textstyle{1\over2}}+n\bigr)\over\Gamma\bigl({\textstyle{1\over2}}+m\bigr)} {n\choose m}
=2^{-2n} {(2n)!\over(2m)!\,(n-m)!} &(A.12)\cr
\noalign{\smallskip}
b_{2m+1,2n+1} &=2^{-(2m+1)}
{\Gamma\bigl({\textstyle{3\over2}}+n\bigr)\over\Gamma\bigl({\textstyle{3\over2}}+m\bigr)} {n\choose m}
=2^{-(2n+1)} {(2n+1)!\over(2m+1)!\,(n-m)!} &(A.13)\cr
(2x)^n &=\sum_{m=0}^{[n/2]} {n!\over m!\,(n-2m)!}\,H_{n-2m}\,. &(A.14)\cr
}$$
These expressions of $b_{m,n}$ can also be obtained directly with the following lemma, setting
$\rho_k=(-)^k\,2^{2k}\,\Gamma\bigl({1\over2}+k\bigr)$ and $\sigma_j=\Gamma\bigl({1\over2}+j\bigr)$.

\smallskip
\noindent{\bf Lemma A.}
{\sl
With $\rho_j$ and $\sigma_j$ some non-zero complex numbers, the matrices ${\bf A}$ and ${\bf B}$ defined by,
$A_{j,k}:=(-1)^j {k\choose j} \rho_k/\sigma_j $ and $B_{j,k}:=(-1)^j {k\choose j} \sigma_k/\rho_j$
for $j,k=0,1\ldots$, are inverse one from the other.
}

\smallskip
\noindent{\bf Proof.}
Since the binomial coefficient ${k\choose j}$ vanishes if $k<j$, the matrices ${\bf A}$ and ${\bf B}$ are
triangular with zeros under the diagonal and nonzero diagonal elements.
Then, this known lemma
\note{See, e.g., [16] p 3, going from equation (1.8) to equation (1.91).},
is a direct consequence of the binomial formula.
Indeed, ${\bf AB}$ is also a triangular matrix and one has
$$({\bf AB})_{j,k}={\sigma_k\over\sigma_j} \sum_{\ell=0,1,\ldots} (-1^{\ell-j} {\ell\choose j} {k\choose\ell}
={k\choose j} {\sigma_k\over\sigma_j} \sum_{\ell=0}^{k-j} (-1)^\ell {k-j\choose\ell}=\delta_{j,k}\,. \eqno(A.15)$$

\medskip
\noindent (3) {\it Three term recurrence formula.}
From equations I-(3.55), (A.3) and (A.10), one has for $n=0,1,\ldots$
\note{See, e.g., [4] 10.13 (5).},
$$H_{n+1}=(A_n x+B_n) H_n-C_n H_{n-1}\qquad A_n=2\qquad B_n=0\qquad C_n=2 n\,. \eqno(A.16)$$

\medskip
\noindent (4) {\it Differentiation formulae.}
With $w$ defined by equation (A.2) and $f$ and $g$ some polynomials, one has
$$\eqalignno{
w' &=-2x w &(A.17)\cr
(f\,,\,g') &=\int_{-\infty}^\infty (w f)\,dg=-\int_{-\infty}^\infty d(w f)\,g=-(f'-2x f\,,\,g) &(A.18)\cr
(f''-2x f'\,,\,g) &=-(f'\,,\,g')=(f\,,\,g''-2x g')\,. &(A.19)\cr
}$$
These equations can be used in two ways to generate differentiation formulae for $n=0,1,\ldots$.

\smallskip
\noindent $\bullet$ First method

\smallskip
\noindent (i) As any polynomial of degree $n-1$, $H'_n=\sum_{j=0}^{n-1} \alpha_j H_j$ where, from the
orthogonality relation I-(A.2) and equation (A.18),
$$\alpha_j h_j=(H_j\,,\,H'_n)=-(H'_j-2x H_j\,,\,H_n)\,. \eqno(A.20)$$
Since $H'_j-2x H_j$ is a polynomial of degree $j+1$, the scalar product above, and thus $\alpha_j$, vanishes
for $j=0,\ldots,n-2$.
Computing the coefficient of $x^{n-1}$ in $H'_n$ fixes $\alpha_{n-1}=nk_n/k_{n-1}$.
One gets
\note{See, e.g., [4] 10.13 (14).},
$$H'_n=2n H_{n-1}\,. \eqno(A.21)$$
Following the same steps, first order differentiation formula for $x^p H'_n$, with $p=0,1,\ldots$,
can thereby be obtained.
Thus, for $p=1$, considering the polynomial $x H'_n$ of degree $n$, one finds
$$x H'_n=n H_n+2n(n-1) H_{n-2}\,. \eqno(A.22)$$

\smallskip
\noindent (ii) Similarly, being a polynomial of degree $n$, $H''_n-2x H'_n=\sum_{j=0}^n \alpha_j H_j$ where,
from equation (A.19),
$$\alpha_j h_j=(H_j\,,\,H''_n-2x H'_n)=(H''_j-2x H'_j\,,\,H_n)\,. \eqno(A.23)$$
Since $H''_j-2x H'_j$ is a polynomial of degree $j$, $\alpha_j$ vanishes for $j=0,\ldots,n-1$.
Computing the coefficient of $x^n$ in $H''_n-2x H'_n$ fixes $\alpha_n$.
Thereby, $H_n$ satisfies the {\it second order differential equation} (2.36)
\note{See, e.g., [4] 10.13 (12).},
$$A(x) H''_n+B(x) H'_n+\lambda_n H_n=0\qquad A(x)=1\qquad B(x)=-2x\qquad\lambda_n =2n\,. \eqno(A.24)$$

\noindent $\bullet$ Second method (permutating $n$ and $j$ in the relations above)

\noindent {\it Only this method can be extended to the SBO polynomials as shown in section 2.3.}

\smallskip
\noindent For $j=0,\ldots,n$, the polynomial $H'_j$ of degree $j-1$ is orthogonal to $H_n$,
and with equation (A.18),
$$0=(H_n\,,\,H'_j)=\bigl(H'_n-2x H_n\,,\,H_j\bigr)\,. \eqno(A.25)$$
Hence, the particular polynomial $H'_n-2x H_n$ of degree $n+1$ is orthogonal to the subspace
${\cal P}_{n+1}$, and thus it reads $\alpha_{n+1} H_{n+1}$.
Computing the coefficient of $x^{n+1}$ in $H'_n-2x H_n$ fixes $\alpha_{n+1}$.
One gets
$$H'_n-2x H_n=-H_{n+1}\,. \eqno(A.26)$$
Following the same steps, first order differentiation formula for $x^p H'_n$, with $p=0,1,\ldots$,
can thereby be obtained.
Thus, with $p=1$, considering the polynomial $x H'_j$ of degree $j$, one finds
$$x H'_n=-{1\over2} H_{n+2}+(2x^2-1-n) H_n\,. \eqno(A.27)$$

\smallskip Actually, all these differentiation formulae and the three term recurrence formula are not independent.
Using simple algebra (substitutions and derivations), it can be checked that:

\noindent (i) all these differentiation formulae follow from the first order differentiation formula (A.21)
and the three term recurrence formula (A.16);

\noindent (ii) more generally, among the equations (A.16), (A.21), (A.24) and (A.26), any pair of equations
implies the other two equations.
In particular, the first order differentiation formula, e.g., (A.26), and the second order
differential equation (A.24) yield the recurrence equation (A.16).
{\it This kind of method is used in section 2.4 to find a five term recurrence formula
with respect to the degree for the Hermite SBO polynomials in the special case $w_2:=e^{-2x^2}$;}

\noindent (iii) note that the differentiation formulae (A.22) and (A.27) relate Hermite polynomials having
the same parity.
From both equations, one gets the recurrence formula,
$$H_{n+2}=2(2x^2-2n-1) H_n-4n(n-1) H_{n-2} \eqno(A.28)$$
which can be obtained also iterating the three term recurrence formula (A.16).

\medskip
\noindent(5) {\it Generating function.}
The most common generating function reads
\note{See, e.g., [4] 10.13 (19).},
$$F_z(x):=\sum_{n=0}^\infty {1\over n!}\,H_n(x)\,z^n=e^{2xz-z^2}\,. \eqno(A.29)$$

\medskip
\bigskip
\noindent{\elevenrmb Appendix B. Laguerre polynomials}

\bigskip
\noindent(1) {\it Definition.}
{\it (Generalized) Laguerre polynomials} $L_n^{(\alpha)},\,n=0,1,\ldots$ are defined with the measure and
normalization
\note{See, e.g., [4] 10.12 (1) and (2).
According to [3] 22.2.12, $\alpha$ is written between parentheses to avoid any confusion with a power
index.},
$$\eqalignno{
a:=0\qquad b &:=\infty\qquad w:={\rm e}^{-x} x^\alpha\quad \alpha>-1 &(B.1)\cr
\noalign{\smallskip}
k_n &:={(-1)^n\over n!}\quad n=0,1,\ldots\,. &(B.2)\cr
}$$

\smallskip
\noindent(2) {\it Connection coefficients.}
From equations (A.1) and the Euler integral (see footnote 27), the moment reads
$$c_j=\int_0^\infty dx\,e^{-x} x^{\alpha+j}=\Gamma(\alpha+1+j)\quad j=0,1,\ldots\,. \eqno(B.3)$$
Then, all the determinants which occur in the general formulae I-(A.5)--I-(A.10) can be obtained with lemma D.
For $n=0,1,\ldots$ and $m=0,\ldots,n$, one finds
\note{See, e.g., [4] 10.12 (2) for $h_n$, (7) for $a_{m,n}$, (2) for $r_n$ only (see footnote 47 in paper I)
and (13) for $L_n^{(\alpha)}(0)$.},
$$\eqalignno{
Z_n &=\prod_{j=0}^n j!\,\Gamma(\alpha+1+j) &(B.4)\cr
h_n &={\Gamma(\alpha+1+n)\over n!} &(B.5)\cr
a_{m,n} &={(-1)^m\over n!} {\Gamma(\alpha+1+n)\over\Gamma(\alpha+1+m)} {n\choose m}
={(-1)^m\over m!} {\alpha+n\choose n-m} &(B.6)\cr
\widehat r_n &=-n(\alpha+n)\qquad\widehat s_n={1\over2}\,n(n-1)(\alpha+n)(\alpha+n-1) &(B.7)\cr
\noalign{\smallskip}
L_n^{(\alpha)}(0) &={(\alpha+1)_n\over n!}={\alpha+n\choose n}\,. &(B.8)\cr
}$$
Similarly, the connection coefficient $b_{m,n}$, not given in the standard textbook already quoted,
can be obtained from equations I-(A.10), (D.2) and (C.3).
For $m=0,\ldots,n$ one has [3]
\note{See, [3] table 22.10 where, the values of $b_{m,n}$, and also $a_{m,n}$,
are listed only up to $n=12$, but the general expression (B.9) is not given.},
$$\eqalignno{
b_{m,n} &={k_m^{-1}\over m!} \sum_{\ell=0}^m (-1)^{m-\ell} {m\choose\ell}
 {\Gamma(\alpha+1+n+\ell)\over\Gamma(\alpha+1+\ell)} \cr
&=(-1)^m m!\,{\Gamma(\alpha+1+n)\over\Gamma(\alpha+1+m)} {n\choose m}
=(-1)^m\,n!\,{\alpha+n\choose n-m}\,.&(B.9)
}$$
This value of $b_{m,n}$ can also be obtained directly with lemma A, setting
$\rho_j=\Gamma(\alpha+1+j)/j!$ and $\sigma_j=\Gamma(\alpha+1+j)$.

\smallskip
\noindent(3) {\it Three term recurrence formula.}
From equations I-(3.55), (B.2) and (B.7), one has for $n=0,1,\ldots$
\note{See, e.g., [4] 10.12 (3).},
$$\eqalignno{
L_{n+1}^{(\alpha)} &=(A_n x+B_n) L_n^{(\alpha)}-C_n L_{n-1}^{(\alpha)} \cr
\noalign{\smallskip}
(n+1) A_n &=-1\qquad (n+1) B_n=\alpha+2n+1\qquad (n+1) C_n=\alpha+n\,. &(B.10)\cr
}$$

\smallskip
\noindent(4) {\it Differentiation formulae.}
With $w$ defined by equation (B.1) and $f$ and $g$ some polynomials, one has
$$\eqalignno{
(xw)' &=-(x-\alpha-1)w &(B.11)\cr
(f\,,\,xg') &=\int_0^\infty xwf\,dg=-\int_0^\infty d(xwf)\,g
=-\bigl(xf'-(x-\alpha-1)f\,,\,g\bigr) &(B.12)\cr
\bigl(xf''-(x-\alpha-1)f'\,,\,g\bigr) &=-(f'\,,\,xg')=-(xf'\,,\,g')
=\bigl(f\,,\,xg''-(x-\alpha-1)g'\bigr)\,. &(B.13)
}$$
As for the Hermite polynomials in appendix A 4-, these relations can be used in two ways to generate
differentiation formulae for $n=0,1,\ldots$.

\smallskip
\noindent $\bullet$ First method

\smallskip
\noindent (i) Considering the polynomial $x L_n^{(\alpha)}{}'$ with equation (B.12), one finds
\note{See, e.g., [4] 10.12 (12).},
$$x L_n^{(\alpha)}{}'=n L_n^{(\alpha)}-(\alpha+n) L_{n-1}^{(\alpha)}\,. \eqno(B.14)$$

\smallskip
\noindent (ii) Considering the polynomial $x L_n^{(\alpha)}{}''-(x-\alpha-1) L_n^{(\alpha)}{}'$
with equation (B.13), one gets the {\it second order differential equation} (2.36)
\note{See, e.g., [4] 10.12 (10).},
$$A(x) L_n^{(\alpha)}{}''+B(x) L_n^{(\alpha)}{}'+\lambda_n L_n^{(\alpha)}=0
\qquad A(x)=x\qquad B(x)=\alpha+1-x\qquad\lambda_n =n\,. \eqno(B.15)$$

\noindent $\bullet$ Second method (permutating $n$ and $j$ in the relations above).
{\it Only this method can be extended to the SBO polynomials as shown in section 3.3.}

\smallskip
\noindent For $j=0,\ldots,n-1$, considering the polynomial $x L_j^{(\alpha)}{}'$, one finds
\note{See, e.g., [4] 10.12 (12).},
$$x L_n^{(\alpha)}{}'=(n+1) L_{n+1}^{(\alpha)}+(x-\alpha-1-n) L_n^{(\alpha)}\,. \eqno(B.16)$$

\smallskip As for the Hermite polynomials, it can be checked that:

\noindent (i) all these differentiation formulae follow from the first order differentiation formula (B.14)
and the three term recurrence formula (B.10);

\noindent (ii) more generally, among the equations (B.10) and (B.14)--(B.16), any pair of equations
implies both other equations.
In particular, the first order differentiation formula (B.16) and the second order
differential equation (B.15) yield the recurrence equation (B.10).
{\it This kind of method is used in section 3.4 to find a five term recurrence formula
with respect to the degree for the Laguerre SBO polynomials in the special case $w_2:=e^{-2x} x^\alpha$.}

\medskip
\noindent(5) {\it Generating function.}
The most common generating function reads
\note{See, e.g., [4] 10.12 (17).},
$$F_z^{(\alpha)}(x):=\sum_{n=0}^\infty L_n^{(\alpha)}(x)\,z^n
=(1-z)^{-\alpha-1} \exp{xz\over z-1}\quad|z|<1\,. \eqno(B.17)$$

\medskip
\noindent(6) {\it Interrelations between Hermite and Laguerre polynomials.}
For any polynomials $f$ and $g$, a change of variable in the integral yields
$$\int_{-\infty}^\infty dx\,e^{-x^2} f(x) g(x)=\int_0^\infty dx\,e^{-x} x^{-{1\over2}} f(\sqrt x) g(\sqrt x)
=\int_0^\infty dx\,e^{-x} x^{1\over2} {f(\sqrt x)\over\sqrt x} {g(\sqrt x)\over\sqrt x}\,.
\eqno(B.18)$$
It follows that for $n=0,1,\ldots$
\note{See, e.g., [4] 10.13 (2) and (3).},
$$\eqalignno{
\widehat L_n^{(-{1\over2})}(x) &=\widehat H_{2n}(\sqrt x)
\qquad H_{2n}(x)=(-1)^n 2^{2n} n!\,L_n^{(-{1\over2})}(x^2) &(B.19)\cr
\widehat L_n^{({1\over2})}(x) &={\widehat H_{2n}(\sqrt x)\over\sqrt x}
\qquad H_{2n+1}(x)=(-1)^n 2^{2n+1} n!\,x L_n^{({1\over2})}(x^2)\,. &(B.20)\cr
}$$

\medskip
\bigskip
\noindent{\elevenrmb Appendix C. Pochammer symbol and some related relations}

\bigskip
\noindent Definition and properties of the {\it Pochammer symbol} or {\it rising factorial} $(z)_n$
are detailed in [13, 14]
\note{See, e.g., [13] appendix I or [14] section 2.}.
With $n$ a non-negative integer and $z$ some complex number, one has
\note{See, e.g., [3] 6.1.5, 6.1.21 and 6.1.22.},
$$(z)_n:={\Gamma(z+n)\over\Gamma(z)}=(-1)^n n!\,{-z\choose n}
=z(z+1)\cdots(z+n-1)=(-1)^n (-z-n+1)_n\,. \eqno(C.1)$$
The Pochammer symbol satisfies the {\it binomial formula}, i.e. with $n$ a non-negative integer and $z$ and $w$ some
complex numbers
\note{See, e.g., [14] section 2.6.},
$$(z+w)_n=\sum_{\ell=0}^n {n\choose\ell}(z)_\ell\,(w)_{n-\ell}\,. \eqno(C.2)$$

\bigskip
\noindent{\bf Lemma C.1.}
{\sl
With $j$ and $k$ non-negative integers and $c$ some complex number not zero or a negative integer, one has
$$\sum_{\ell=0}^k (-1)^{k-\ell}\,{k\choose\ell}\,{\Gamma(c+j+\ell)\over\Gamma(c+\ell)}
={\displaystyle k!\,{j\choose k}\,{\Gamma(c+j)\over\Gamma(c+k)}} \eqno(C.3)$$
where, the binomial coefficient ${j\choose k}$ vanishes if $j<k$.
}

\medskip
\noindent{\bf Proof.}
Setting $n=k$, $z=c+j$ and $w=-c-k+1$ in the binomial formula (C.2) yields
$$(j-k+1)_k=\sum_{\ell=0}^k {k\choose\ell}\,(c+j)_\ell\,(-c-k+1)_{k-\ell}\,. \eqno(C.4)$$
Then, it follows from equation (C.1) that,
$$(j-k+1)_k=k!\,{j\choose k}
\qquad(c+j)_\ell={\Gamma(c+j+\ell)\over\Gamma(c+j)}
\qquad(-c-k+1)_{k-\ell}=(-1)^{k-\ell} {\Gamma(c+k)\over\Gamma(c+\ell)} \eqno(C.5)$$
thereby recovering equation (C.3) provided $c+j\ne0,-1,\ldots$.
This identity can be proved also directly by recurrence on $k$.

\bigskip
\noindent{\bf Lemma C.2.}
{\sl
With $c$, $z$ and $w$ some complex numbers, $c$ not zero or a negative integer,
and $j$ and $k$ non-negative integers, let,
$$\eqalignno{
S_1(j,k,c;z,w) &:=\sum_{p=0}^{{\rm min}(j,k)} {j\choose p}\,{k\choose p}
\,{p!\over\Gamma(c+p)}\,(zw)^p\,(1+z)^{j-p}\,(1+w)^{k-p} &(C.6)\cr
S_2(j,k,c;z,w) &:=\sum_{p=0}^j \sum_{q=0}^k {j\choose p}\,{k\choose q}
\,{\Gamma(c+p+q)\over\Gamma(c+p)\,\Gamma(c+q)}\,z^p\,w^q &(C.7)\cr
S(j,k,c;z,w) &:={(1+z)^j\,(1+w)^k\over\Gamma(c)}\,{\rm F}\left(-j,-k;c;{zw\over(1+z)\,(1+w)}\right) &(C.8)\cr
}$$
where, ${\rm F}$ is an hypergeometric function.
Then, one has
$$S(j,k,c;z,w)=S(k,j,c;w,z)=S_1(j,k,c;z,w)=S_2(j,k,c;z,w)\,. \eqno(C.9)$$
}

\medskip
\noindent{\bf Proof.}
From equations (C.1), in terms of the Pochammer symbol, one has
$$\eqalignno{
S_1(j,k,c;z,w) &={(1+z)^j\,(1+w)^k\over\Gamma(c)}\,\sum_{p=0}^{{\rm min}(j,k)} {(-j)_p\,(-k)_p\over(c)_p}
\,{1\over p!}\,\left({zw\over(1+z)\,(1+w)}\right)^p \cr
&=S(j,k,c;z,w) &(C.10)\cr
\noalign{\smallskip}
S_2(j,k,c;z,w) &={1\over\Gamma(c)}\,\sum_{p=0}^j\sum_{q=0}^k {(-j)_p\,(-k)_q\,(c)_{p+q}\over(c)_p\,(c)_q}
\,{(-z)^p\over p!}\,{(-w)^q\over q!} \cr
&={1\over\Gamma(c)}\,{\rm F}_2(c,-j,-k,c,c;-z,-w) &(C.11)
}$$
where, ${\rm F}_2$ is a special case of a generalized hypergeometric series of two variables listed by Horn
and reducible to an hypergeometric function [12]
\note{See, e.g., [12] 5.7.1 (7) and 5.10 (3).},
$${\rm F}_2(c,-j,-k,c,c;-z,-w)=(1+z)^j\,(1+w)^k\,{\rm F}\left(-j,-k;c;{zw\over(1+z)\,(1+w)}\right)\,. \eqno(C.12)$$
The symmetry property $(j,z)\leftrightarrow(k,w)$ is obvious.
This completes the proof of equation (C.9).

\bigskip
\noindent{\bf Corollary C.}
{\sl
Special values of $S(j,k,c;z,w)$, see lemma C.2 equations (C.6)--(C.9):

\smallskip
$$\eqalignno{
S(j,k,c;z,-1) &={j\choose k} {k!\over\Gamma(c+k)} (-z)^k (1+z)^{j-k} &(C.13)\cr
S(j,k,c;-1,-1) &={j!\over\Gamma(c+j)} \delta_{j,k} &(C.14)\cr
S(j,k,c;z,-z-1) &=(1+z)^j (-z)^k {\Gamma(c+j+k)\over\Gamma(c+j) \Gamma(c+k)} &(C.15)\cr
}$$
where, in equation (C.13), the binomial coefficient ${j\choose k}$ vanishes if $j<k$.
}

\medskip
\noindent{\bf Proof.}
For $z=-1$ or $w=-1$, the first two equations follow directly from lemma C.2.
Note that $S_1$ and $S_2$ have no singularity, in particular at $z=-1$ or/and $w=-1$.
If $w=-1$, only the term $p=k$ contributes to equation (C.6), giving equation (C.13).
Alternately, setting $w=-1$ in equation (C.7), the sum over $q$ can be performed with lemma C.1.
For $z=-w-1$, one has $zw/((1+z)\,(1+w))=1$, then from the {\it Gauss summation theorem}
\note{See, e.g., [3] 15.1.20 or [13] (1.7.6).},
$${\rm F}(-j,-k,c;1)={\Gamma(c) \Gamma(c+j+k)\over\Gamma(c+j) \Gamma(c+k)} \eqno(C.16)$$
ending the proof of equation (C.15).

\medskip
\bigskip
\noindent{\elevenrmb Appendix D. Some determinants with gamma functions as entries}

\bigskip
\noindent{\bf Lemma D.}
{\sl
With $n$ a positive integer and $c$ some complex number not zero or a negative integer, let ${\bf M}$ be an
$n\times n$ matrix which element $M_{j,k}$ is equal to $\Gamma(c+j+k)$, except possibly for the last row,
$${\bf M}:=\pmatrix{
\bigl(\Gamma(c+j+k)\bigr)_{j=0,\ldots,n-2\hfill\atop k=0,\ldots,n-1\hfill} \cr
\noalign{\smallskip}
(M_{n-1,k})_{k=0,\ldots,n-1} \cr
}\,. \eqno(D.1)$$
Then, one has
$$\eqalignno{
{\det{\bf M}\over\prod_{j=0}^{n-2}j!\,\Gamma(c+j)} &=\Gamma(c+n-1)
\sum_{\ell=0}^{n-1}(-1)^{n-1-\ell} {n-1\choose\ell} {1\over\Gamma(c+\ell)}\,M_{n-1,\ell} &(D.2)\cr
\det\bigl[\Gamma(c+j+k)\bigr]_{j,k=0,\ldots,n-1} &=\prod_{j=0}^{n-1}j!\,\Gamma(c+j)\,. &(D.3)\cr
}$$
}

\medskip
\noindent{\bf Proof.}
A determinant is not changed if one adds to any column a linear combination of other columns.
Replacing for $k=0,\ldots,n-1$ the column ${\cal C}_k$ of $\det{\bf M}$ by,
$${\cal C}_k+\sum_{\ell=0}^{k-1} (-1)^{k-\ell} {k\choose\ell} {\Gamma(c+k)\over\Gamma(c+\ell)} {\cal C}_\ell
=\sum_{\ell=0}^k (-1)^{k-\ell} {k\choose\ell} {\Gamma(c+k)\over\Gamma(c+\ell)} {\cal C}_\ell \eqno(D.4)$$
yields a new matrix ${\bf M'}$ with an equal determinant and which is triangular.
Indeed, from definition of ${\bf M}$ and lemma C.1, one has for $j=0,\ldots,n-2$ and $k=0,\ldots,n-1$,
$$M_{j,k}'=\Gamma(c+k) \sum_{\ell=0}^k (-1)^{k-\ell} {k\choose\ell}{\Gamma(c+j+\ell)\over\Gamma(c+\ell)}
=k!\,{j\choose k} \Gamma(c+j) \eqno(D.5)$$
which vanishes for $j<k$.
Then, the determinant is the product of the diagonal elements $j!\,\Gamma(c+j)$ for $j=0,\ldots,n-2$ and
$M_{n-1,n-1}'$.
From equation (D.4), this last element reads
$$M_{n-1,n-1}'=\Gamma(c+n-1) \sum_{\ell=0}^{n-1} (-1)^{n-1-\ell} {n-1\choose\ell}
{1\over\Gamma(c+\ell)} M_{n-1,\ell} \eqno(D.6)$$
completing the proof of equation (D.2).
If moreover, $M_{n-1,k}=\Gamma(c+n-1+k)$, using again the equation (D.5) for $j=k=n-1$ yields the result (D.3)
already published in [15, 14]
\note{See [15] equation (A.12) or [14] equation (4.5).}.

\medskip
\bigskip
\noindent{\elevenrmb Appendix E. Special cases of Hermite standard bloc orthogonal polynomials}

\bigskip
\noindent With the weight functions, $x\in(-\infty,\infty),\ w:=e^{-x^2}$ and $w_2:=e^{-2x^2}$, one has:

\smallskip
\noindent $i=0\qquad n=1,2\ldots$
$$\eqalignno{
\widehat P_{0;1} &=x &(E.1)\cr
\widehat P_{0;2} &=x^2-{1\over4} &(E.2)\cr
\widehat P_{0;3} &=x^3-{3\over4}\,x &(E.3)\cr
\widehat P_{0;4} &=x^4-{3\over2}\,x^2+{3\over16} &(E.4)\cr
\widehat P_{0;5} &=x^5-{5\over2}\,x^3+{15\over16}\,x\,; &(E.5)\cr
}$$
\noindent $i=1\qquad n=1,2,\ldots$
\note{In the case $i=1$, `Hermite polynomials under a constraint of zero average' have already been
considered in [6] and [8], with different choices of weight functions and normalization:
\vskip 0mm
\noindent - in [6] (1) and appendix A, the polynomials $A_n:=2^n\,x^n+{\rm O}(x^{n-1}),\,n=1,2,\ldots$,
with $w:=e^{-{1\over2}x^2}$ and $w_2:=e^{-x^2}$.
From equation (E.12) below with $c={1\over2}$, one gets
$A_n(x)=2^{{3\over2}n}\,\widehat P_{1;n}(x/\sqrt2),\,n=1,2,\ldots\,;$
\vskip 0mm
\noindent - in [8] (43), the polynomials $Q_{2n}:=2^{n-1}\,x^n+{\rm O}(x^{n-1}),\,n=1,2,\ldots$,
with $w:=e^{-{1\over2}x^2}$ and $w_2:=e^{-x^2}$.
Then, one has $Q_{2n}=2^{-(n+1)}\,A_{2n},\,n=1,2,\ldots$.
Note that with this choice of $w_2$, one has $\widehat Q_{2;n}=\widehat H_n,\,n=0,1,\ldots$;
\vskip 0mm
\noindent - in [6] section 4, the polynomials $G_n:=c_n\,\widehat P_{1;n},\,n=1,2,\ldots$, with
$(G_m\,,\,G_n)_2=\sqrt{\pi/2}\,\delta_{m,n}$, $w:=e^{-x^2}$ and $w_2:=e^{-2x^2}$.
From equations (2.18) and (2.19), one finds
$c_{2n+1}=\pm\,2^{2n+1}/\sqrt{(2n+1)!},\,n=0,1,\ldots$ and
$c_{2n}=\pm\,2^{2n+1}\,\sqrt{2n}/\sqrt{(2n+1)!},\,n=1,2,\ldots$;
\vskip 0mm
\noindent - in [8] (44), (45a)--(45d), the polynomials $\Gamma_n:=|c_n|\,\widehat P_{1;n},\,n=1,2,\ldots$,
with $c_n$ as for $G_n$ above, $(\Gamma_m\,,\,\Gamma_n)_2=\sqrt{\pi/2}\,\delta_{m,n}$, $w:=e^{-x^2}$ and
$w_2:=e^{-2x^2}$, and in addition, with a positive coefficient of $x^n$ in $\Gamma_n$.
},
$$\eqalignno{
\widehat P_{1;2} &=x^2-{1\over2} &(E.6)\cr
\widehat P_{1;4} &=x^4-{7\over4}\,x^2+{1\over8} &(E.7)\cr
\widehat P_{1;6} &=x^6-4\,x^4+{47\over16}\,x^2-{11\over32}\,; &(E.8)\cr
}$$
\noindent $i=2\qquad n=2,3,\ldots$
\note{The `Hermite polynomials constrained by a zero momentum', considered in [6] section 3 (17),
with $w:=e^{-{1\over2}x^2}$ and $w_2:=e^{-x^2}$, {\it is not the case} $i=2$.
Indeed, the one-dimensional subspace ${\cal E}_1$ considered there, and spanned by $\{x^1\}$,
is not ${\cal P}_2$ spanned by $\{x^0,x^1\}$.
Nevertheless, as discussed in the conclusion I-4 (see footnote 44 in paper I),
from parity arguments, it follows that in this
non-standard case, the block orthogonal polynomials $\widehat P_n,\,n=0,2,3,\ldots$ are given by,
$\widehat P_{2n}=\widehat Q_{2;2n},\,n=0,1,\ldots$ and $\widehat P_{2n+1}=\widehat P_{2;2n+1},\,n=0,1,\ldots$.
From equation (E.12) with $c={1\over2}$, one gets with $B_n:=2^n\,x^n+{\rm O}(x^{n-1}),\,n=0,2,3,\ldots$,
$B_{2n}=H_{2n},\,n=0,1,\ldots$
and $B_{2n+1}(x)=2^{3(n+{1\over2})}\,\widehat P_{2;2n+1}(x/\sqrt2),\,n=1,2,\ldots$.},
$$\eqalignno{
\widehat P_{2;3} &=x^3-{3\over2}\,x &(E.9)\cr
\widehat P_{2;5} &=x^5-{13\over4}\,x^3+{9\over8}\,x\,. &(E.10)\cr
}$$

If, instead of $w$ and $w_2$, one considers the following weight functions,
$$w^{(c)}(x):=w(\sqrt c x)=e^{-cx^2}
\qquad w_2^{(c)}:=w_2(\sqrt c x)=e^{-2cx^2}\quad c>0 \eqno(E.11)$$
then, changing $x$ into $\sqrt c x$ in the integral defining the scalar product $(\,,\,)_2$,
the corresponding monic Hermite SBO polynomials $\widehat P_{i;n}^{(c)}$ and the orthogonality constant
$\widehat H_{i;n}^{(c)}$ read
$$\widehat P_{i;n}^{(c)}(x)=c^{-{1\over2}n} \widehat P_{i;n}(\sqrt c x)
\qquad\widehat H_{i;n}^{(c)}=c^{-({1\over2}+n)} \widehat H_{i;n}\,. \eqno(E.12)$$

\medskip
\bigskip
\noindent{\elevenrmb Appendix F. Special cases of Laguerre standard bloc orthogonal polynomials}

\bigskip
\noindent With the weight functions, $x\in[0,\infty),\ w:=e^{-x} x^\alpha$ and $w_2:=e^{-2x} x^\alpha,\ \alpha>-1$,
one has:

\smallskip
\noindent $i=0\qquad n=1,2,\ldots$
$$\eqalignno{
\widehat P_{0;1} &=x-{1\over2}\,(\alpha+1) &(F.1)\cr
\widehat P_{0;2} &=x^2-(\alpha+2)\,x+{1\over4}\,(\alpha+1)(\alpha+2) &(F.2)\cr
\widehat P_{0;3} &=x^3-{3\over2}\,(\alpha+3)\,x^2+{3\over4}\,(\alpha+2)(\alpha+3)\,x
-{1\over8}\,(\alpha+1)(\alpha+2)(\alpha+3) &(F.3)\cr
\widehat P_{0;4} &=x^4-2\,(\alpha+4)\,x^3+{3\over2}\,(\alpha+3)(\alpha+4)\,x^2
-{1\over2}\,(\alpha+2)(\alpha+3)(\alpha+4)\,x \cr
&\phantom{=}+{1\over16}\,(\alpha+1)(\alpha+2)(\alpha+3)(\alpha+4)\,; &(F.4)\cr
}$$

\smallskip
\noindent $i=1\qquad n=1,2,\ldots$
\note{In the case $i=1$, `modification of Laguerre polynomials by a constraint of zero average' have already been
considered in [7] section 2, with different choices of weight functions and normalization:
the monic polynomials $G_n^d,\,n=1,2,\ldots$, with $\alpha:=d-1$, $w:=e^{-{1\over2}x}\,x^{d-1}$
and $w_2:=e^{-x}\,x^{d-1}$.
Setting $c={1\over2}$ in equation (F.16) below yields:
$G_n^d(x)=2^n\,\widehat P_{1;n}(x/2)|_{\alpha=d-1},\,n=1,2,\ldots$.
Note that with this choice of $w_2$, $\widehat Q_{2;n}=\widehat L_n^{(\alpha)},\,n=0,1,\ldots$.},
$$\eqalignno{
\widehat P_{1;1} &=x-(\alpha+1) &(F.5)\cr
\widehat P_{1;2} &=x^2-{1\over2}\,(3\alpha+5)\,x+{1\over2}\,(\alpha+1)^2 &(F.6)\cr
\widehat P_{1;3} &=x^3-(2\alpha+5)\,x^2+{1\over4}\,(5\alpha^2+19\alpha+20)\,x
-{1\over4}\,(\alpha+1)(\alpha^2+3\alpha+4) &(F.7)\cr
\widehat P_{1;4} &=x^4-{1\over2}\,(5\alpha+17)\,x^3+{3\over4}\,(3\alpha^2+17\alpha+26)\,x^2 \cr
&\phantom{=}-{1\over8}\,(7\alpha^3+48\alpha^2+125\alpha+108)\,x
+{1\over8}\,(\alpha+1)^2(\alpha^2+5\alpha+12) &(F.8)\cr
\widehat P_{1;5} &=x^5-(3\alpha+13)\,x^4+{1\over2}\,(7\alpha^2+53\alpha+106)\,x^3
-{1\over2}\,(4\alpha^3+39\alpha^2+137\alpha+162)\,x^2 \cr
&\phantom{=}+{1\over16}\,(9\alpha^4+98\alpha^3+447\alpha^2+886\alpha+648)\,x \cr
&\phantom{=}-{1\over16}\,(\alpha+1)(\alpha^4+10\alpha^3+47\alpha^2+86\alpha+72)\,; &(F.9)\cr
}$$

\smallskip
\noindent $i=2\qquad n=2,3,\ldots$
$$\eqalignno{
\widehat P_{2;2} &=x^2-2(\alpha+2)\,x+(\alpha+1)(\alpha+2) &(F.10)\cr
\widehat P_{2;3} &=x^3-{1\over2}(5\alpha+13)\,x^2+2(\alpha+2)^2\,x-{1\over2}(\alpha+1)^2(\alpha+2) &(F.11)\cr
\widehat P_{2;4} &=x^4-(3\alpha+10)\,x^3+{1\over4}(13\alpha^2+71\alpha+102)\,x^2 \cr
&\phantom{=}-{1\over2}(\alpha+2)(3\alpha^2+13\alpha+18)\,x+{1\over4}(\alpha+1)(\alpha+2)(\alpha^2+3\alpha+6)
&(F.12)\cr
\widehat P_{2;5} &=x^5-{1\over2}(7\alpha+29)\,x^4+{1\over4}(19\alpha^2+135\alpha+254)\,x^3 \cr
&\phantom{=}-{1\over8}(25\alpha^3+222\alpha^2+719\alpha+810)\,x^2
+{1\over2}(\alpha+2)^2(2\alpha^2+11\alpha+27)\,x \cr
&\phantom{=}-{1\over8}(\alpha+1)^2(\alpha+2)(\alpha^2+5\alpha+18) &(F.13)\cr
\widehat P_{2;6} &=x^6-4(\alpha+5)\,x^5+{1\over2}(13\alpha^2+115\alpha+268)\,x^4
-{1\over2}(11\alpha^3+127\alpha^2+526\alpha+752)\,x^3 \cr
&\phantom{=}+{1\over16}(41\alpha^4+538\alpha^3+2947\alpha^2+7418\alpha+7056)\,x^2 \cr
&\phantom{=}-{1\over8}(\alpha+2)(5\alpha^4+58\alpha^3+319\alpha^2+770\alpha+720)\,x \cr
&\phantom{=}+{1\over16}(\alpha+1)(\alpha+2)(\alpha^4+10\alpha^3+59\alpha^2+122\alpha+144)\,. &(F.14)\cr
}$$

If, instead of $w$ and $w_2$, one considers the following weight functions,
$$w^{(c)}(x):=c^{-\alpha} w(cx)=e^{-cx} x^\alpha
\qquad w_2^{(c)}(x):=c^{-\alpha} w_2(cx)=e^{-2cx} x^\alpha\quad c>0 \eqno(F.15)$$
then, changing $x$ into $cx$ in the integral defining the scalar product $(\,,\,)_2$,
the corresponding monic Laguerre SBO polynomials $\widehat P_{i;n}^{(c)}$ and the orthogonality constant
$\widehat H_{i;n}^{(c)}$ read
$$\widehat P_{i;n}^{(c)}(x)=c^{-n} \widehat P_{i;n}(c\,x)
\qquad\widehat H_{i;n}^{(c)}=c^{-(\alpha+1+2n)} \widehat H_{i;n}\,. \eqno(F.16)$$

\medskip
\bigskip
\noindent {\elevenrmb References}

\bigskip
\item{[1]} Normand J-M 2007
Block orthogonal polynomials: I. Definitions and properties
{\it J. Phys. A: Math. Theor.} {\bf 40} 2341--69, {\it matph-ph}/0606036

\item{[2]} Szeg\"o G 1975
{\it Orthogonal Polynomials} Colloquium Publications Vol. XXIII
4th edn
(New York: American Mathematical Society)

\item{[3]} Abramowitz M and Stegun I A 1972
{\it Handbook of Mathematical Functions}
(New York: Dover)

\item{[4]} Bateman H 1953
{\it Higher Transcendental Functions} vol 2
(New York: McGraw-Hill)

\item{[5]} Gradshteyn I S and Ryzhik I M 2000
{\it Table of Integrals, Series, and Products} 6th edn
(New York: Academic)

\item{[6]} Giraud B G, Mehta M L and Weiguny A 2004
Orthogonal polynomials sets with finite codimensions
{\it C. R. Physique} {\bf 5} 781--7

\item{[7]} Giraud B G 2005
Constrained orthogonal polynomials
{\it J. Phys. A: Math. Gen.} {\bf 38} 7299--311

\item{[8]} Giraud B G, Weiguny A, Wilets L 2005
Coordinates, modes and maps for the density functional
{\it Nucl. Phys. A } {\bf 761} 22--40

\item{[9]} Hohenberg P and Kohn W 1964
Inhomogeneous electron gas
{\it Phys. Rev.} {\bf 136} B864--71

\item{[10]} Prudnikov A P, Brychkov Yu A and Marichev O I 1992
{\it Integrals and Series} vol 2
(New York: Gordon and Breach)

\item{[11]} Andrews G E, Askey R and Roy R 1999
{\it Special Functions (Encyclopedia of Mathematics and its Applications vol 71)}
(Cambridge: Cambridge University Press)

\item{[12]} Bateman H 1953
{\it Higher Transcendental Functions} vol 1
(New York: McGraw-Hill)

\item{[13]} Slater L J 1966
{\it Generalized Hypergeometric Functions}
(London: Cambridge University Press)

\item{[14]} Normand J-M 2004
Calculation of some determinants using the $s$-shifted factorial
{\it J. Phys. A: Math. Gen.} {\bf 37} 5737--62

\item{[15]} Mehta M L and Normand J-M 1998
Probability density of the determinant of a random Hermitian matrix
{\it J. Phys. A: Math. Gen.} {\bf31} 5377--91

\item{[16]} Rademacher H 1973
{\it Topics in Analytic Number Theory}
(Berlin: Springer)

\end